# Managing Financial Climate Risk in Banking Services: A Review of Current Practices and the Challenges Ahead

## Víctor Cárdenas*

*Institute for Resources, Environment and Sustainability, University of British Columbia

**Abstract-**. The document discusses the financial climate risk in the context of the banking industry, emphasizing the need for a comprehensive understanding of climate change across different spatial and temporal scales. It highlights the challenges in estimating physical and transition risks, specifically extreme events and limitations of current climate models. The document also reviews current gaps in assessing physical and transition risks, including the development, improvement of modeling frameworks, highlighting the need for detailed databases of exposed physical assets and climatic hazard modeling. It also emphasizes the importance of integrating financial climate risks into financial risk management practices, particularly in smaller banks and lending organizations.

**JEL Codes-** G10, G30, Q54, Q56.

## I. INTRODUCTION

The aim of this paper is to outline financial risks that the banking industry (BIS, 2024b) has traditionally evaluated and used to establish a framework to introduce new risks related to climate change. The objective is to assess the financial climate risks, methodological developments, and potential expansion opportunities.

Climate change does not go unnoticed by the financial sector (Carney Mark, 2015), which includes the banking industry; this industry is frequently an integral component of the supply chain that directly or indirectly emit greenhouse gases. By providing financing, the banking industry contributes to the enormous and complex problem that climate change inherently presents. For example, in Canada, according to an Oxfam report (2022), the financed emissions through the Canadian banking system are equivalent to roughly 77% (RBC) and 95% (Laurentian) of the total assets of each of the eight banks, or nearly C$5.670 trillion of the C$6.934 trillion reported on 2020 balance sheets, 82%. According to the report, if the financing from the top eight banks were the GDP of a sovereign country, they would have counted as the fifth largest emitter of greenhouse gases in the world at the end of 2020 in terms of asset-backed emissions, behind China, the United States, India and Russia.

The banking industry is itself crucial for any economy (Apostolik et al., 2009) by providing access to the payment systems within and among nations. The operational efficiency of industries directly linked to greenhouse gas emissions, as well as that of every other sector of the economy, can be influenced by public policy measures that mitigate emissions.

These measures could directly or indirectly affect the operation and financial performance of the banking industry and consequently the financial system and economic growth. Therefore, it is critical to qualitatively and quantitatively comprehend the industry's level of exposure to financial climate risk across all its facets, ranging from significant to minor (Szulczyk Kenneth, 2014).

This paper provides an analysis of financial climate risk estimation, focusing on both physical and transition risks, particularly in the context of the banking industry. It discusses the assessment of financial risk associated with the physical impacts of climate change, emphasizing the need for knowledge of climate change across multiple spatial and temporal scales to assess these risks. The document also mentions the challenges in assessing physical risk, such as the difficulty in estimating extreme events and the limitations of current climate models in providing detailed and reliable information at smaller spatial and temporal scales.

Regarding physical risk, the document focuses on the coverage of physical climate hazard models. It also highlights the gaps in the guidance, such as the absence of a shared robust understanding and approach to identifying and assessing physical climate risks. The document also discusses the literature on transition risk and describes the evolution in the types of models for assessing it, including main strengths and weaknesses.

The document examines the assessment and management of financial climate risk, addressing physical and transition risks, and emphasizing the need for improved methodologies and data quantity and quality to effectively manage these risks within the banking industry. It also highlights the importance of understanding the interconnections between physical risk and economic sectors, as well as the interconnections with the rest of the financial industry. The document concludes by acknowledging the rapid growth in the literature on financial climate risk, particularly in terms of transition risk and physical risk, while also recognizing the remaining challenges in this area. As part of the analysis, one of the challenges for both risks is in interpreting estimations of physical and transition risk in the context of risk management methodologies within banks, especially in small banks.

The document is integrated as follows, there are 5 sections in addition to the introduction. The second section discusses current risk management practices in banks. The third section deals with climate risk, physical risk, and transition risk, respectively. Finally, the fourth section presents the conclusions and the fifth is the references.

## II. CURRENT RISK MANAGEMENT PRACTICES IN THE BANKING BUSINESS

Banks are entities that, by their very nature, provide a service that has various repercussions with the economy as a whole, with savers and borrowers. The example that historically set a precedent was the 1929 crisis in the United States, followed by the crisis in the stock market, which had important repercussions on the health of banks, and escalated into a deep recession in the U.S. economy.

Banks have access to the country's payment system, in addition to being depositories of people's savings, as well as channeling savings in the form of investments to productive activities in the economy. When one of the banks defaults on its payments, whether to its customers or to its bondholders, it is not only a risk for the bank's shareholders, but also a risk for the economy due to the enormous risk of a bank run, that is, the risk that all the banks will be demanded the total savings of the savers, which in principle are not available, since they are invested (Szulczyk Kenneth, 2014).

The risk of bank runs has been the main driver of concern for regulators around the world and historically regulations have been written with this in mind (Ishtiaq, 2015). In 1974, one of the global guidelines in banking regulation emerged, the 10 most developed economies, through the central banks, created the Basel Committee on Banking Supervision. The committee agreed to discuss the possibility of developing minimum capital, liquidity, and funding standards. In 1988 a first body of standards emerged, later to be known by the name of the committee's host city, Basel, the first version of the proposed standards was named Basel I, followed in 2004 by Basel II, and in 2010 Basel III was issued (De Bandt et al., 2023). Although the standards were non-binding, they laid the groundwork for a standardized rule for all adhering countries.

Currently, Basel III provides strong risk management recommendations throughout its framework. By suggesting stricter capital, liquidity, and leverage requirements, it incentivizes banks to adopt more robust risk management practices to identify, measure, monitor, and control the main financial risk of the banking industry (BIS, 2024b).

In the international regulatory frameworks, the banking industry is commonly highlighted as a source of uncertainty for the banking business with the following key risks: credit risk, market risk, liquidity risk, and operational risk, which are described below.

**Table 2. Definition of financial risk for the banking industry**

| Type of risk | Definition |
|---|---|
| Credit risk | Potential loss that a financial institution may incur due to the failure of a borrower to meet their financial obligations. This includes the risk of default on loans or other credit agreements. |
| Market risk | Potential financial loss due to adverse movements in market prices, including foreign exchange, traded debt securities, equities, commodities, and derivatives. |
| Liquidity risk | Potential challenges in meeting its short-term obligations due to a lack of liquid assets to cover deposit withdrawals. This risk arises from the difference in duration between loans (assets) and deposits, as banks typically hold only a fraction of the deposits. |
| Operational risk | Potential loss resulting from inadequate or failed internal processes, systems, people, or external events. This includes risks associated with information technology, data management, human error, and fraud. |

The evolution of the Basel Accords was gradual in terms of the risks it regulated (BIS, 2024a). It started with credit risk as a priority, but later on, liquidity, market, and operational risk were taken into account. The following table shows its evolution.

**Table 1. Types of financial risk for banking institutions according to the Basel framework**

| Risk Type | Basel I | Basel II | Basel III |
|---|---|---|---|
| Credit Risk | Only considered credit risk in capital calculations, using a single risk weight for all assets. | Introduced risk-sensitive capital requirements based on credit ratings, allowing banks to assign different weights to assets based on their credit risk. | Enhanced credit risk management with more sophisticated approaches for measuring and managing credit risk, including stress testing and forward-looking provisioning. |
| Liquidity Risk | Lacked explicit liquidity risk regulations. | Addressed liquidity risk more comprehensively, introducing liquidity coverage ratio (LCR) and net stable funding ratio (NSFR) to ensure banks maintain sufficient liquidity buffers. | Further strengthened liquidity risk management by refining liquidity standards, including additional requirements for liquid assets and introducing the liquidity coverage ratio (LCR) and net stable funding ratio (NSFR). |
| Market Risk | Basel I didn't include specific market risk regulations. | Basel II introduced a standardized approach and internal models approach for calculating market risk capital requirements, considering factors like value-at-risk (VaR). | Basel III enhanced market risk regulations by introducing standardized and internal model approaches, incorporating more stringent capital requirements, and expanding the scope of instruments covered. |
| Operational Risk | Basel I didn't address operational risk explicitly. | Basel II introduced operational risk as a separate category with capital requirements based on historical loss data or statistical models. | Basel III enhanced operational risk management by introducing a more sophisticated approach to capital requirements, including the Advanced Measurement Approach (AMA) for larger banks. |

Source: taken from https://www.bis.org/bcbs/history.htm

In the latest release of the Accord, Basel III incorporated several measures with a special focus on recognizing the global crisis in the structured bond market, so one of the main contributions of this regulatory framework versus previous ones was the incorporation of risk management measures, although the framework was not alien to risk management, in this version of the standard explicit risk management methodologies applied to the banking business were added (BIS, 2024b).

In this way, the banking business, after the 2007/2008 US subprime mortgage crisis, implicitly recognized the solvency of financial institutions and introduced the concept of financial stability. Recognizing that there are financial institutions that due to their characteristics (size and role in the economy) simply cannot fail (too big to fail), which could be relevant for a single economy, but for multiple economies or globally systemically important banks.

Although the international standard suggests following a risk-based approach, as suggested by Basel III, not all countries in the world follow it, some follow parts of the standard, and a few comply fully. For example, the United States does not follow it fully, this fact has again been discussed because the recent crisis in 2023 of a Californian niche-bank (Silicon Valley Bank) showed that the system still has areas of opportunity, precisely because the bank's failure was a focused risk management practice, which was omitted or ignored by the bank and by the regulatory authority as well.

II.1. CONVENTIONAL MODELING APPROACH BY TYPE OF RISK

II.1.1. CREDIT RISK

Credit risk, or the possibility that money owed will not be returned. This kind of risk has been present in financial, trade, and commerce transactions since the dawn of civilization (Baesens & van Gestel, 2009a). Credit risk management has become increasingly crucial because of the numerous failures throughout history and the resulting impacts on the economy and society.

Identification of potential hazards, evaluation of these risks, application of suitable remedies, and consequent use of risk models are all steps in the credit risk management process. Over the past 60 years, consumer credit has increased dramatically, and efficient credit risk management technologies have been essential to this growth (Van-Deventer Donarld R. et al., 2013).

Two distinct categories of methodologies exist for estimating credit risk: those that examine borrowers individually and those that employ a portfolio approach.

For the former, credit scoring and credit rating methodologies are implemented. The methodology is consistently applied on an individual basis to a borrower who is seeking to ascertain its solvency to finance the repayment of a loan or set of loans, within the framework of its financial management and business (Skoglund Jimmy & Chen Wei, 2015).

The portfolio borrower methodology, conversely, conducts an individual analysis within the framework of a portfolio. Where analysis of default or non-default occurs of an individual loan that belongs to a portfolio. A key component of this framework is portfolio diversification; therefore, the probability of incurring such a substantial loss is considerably diminished for a portfolio consisting of hundreds of loans, as the probability that all loans default simultaneously is numerous times smaller than the probability that a single loan defaults (Gordy, 2000).

The modeling of portfolio credit risk is based on the estimation of the following probabilities: Probability of Default (PD), Loss Given Default (LGD), Exposure at Default (EAD). Currently, one of the main methodological approaches is the structural model. The structural approach aims to provide an explicit relationship between default risk and the capital structure of the firm, while the reduced form approach models credit defaults as exogenous events driven by a stochastic process (such as a Poisson jump process) (Hurd Tom, 2010).

**Table 3. Key probabilities for structural model on credit risk models for the banking industry**

| Aspect | Probability of Default (PD) | Loss Given Default (LGD) | Exposure at Default (EAD) |
|---|---|---|---|
| Definition | Likelihood of a borrower defaulting on a loan | Share of an asset value lost if a borrower defaults | Loss exposure value at the time of default |
| Calculation | Based on historical default rates, borrower characteristics, and macroeconomic indicators | Historical recovery rates, collateral values, and contractual terms | Loan balance at the time of default, considering future draws and changes in risk factors |
| Purpose | Assess the likelihood of credit default | Estimate potential credit losses | Determine the magnitude of potential loss |
| Importance | Essential for assessing creditworthiness of borrowers and setting risk premiums | Crucial for estimating credit losses and calculating expected losses | Critical for determining the capital reserves needed to cover potential losses |
| Models | PD estimation include logistic regression, decision trees, and learning algorithms | LGD models utilize linear regression, machine learning, and historical data analysis | EAD models may employ linear regression, Monte Carlo simulation, or banking risk management techniques |
| Output | Probability value ranging from 0 to 1 | Percentage value ranging from 0% to 100% | Monetary value representing the exposure amount |
| Application | Used in credit scoring, credit rating, and loan pricing | Incorporated into risk management strategies, loan pricing, and portfolio optimization | Integral to regulatory capital calculations, risk-based pricing, and stress testing |
| Regulatory Compliance | Basel Accords (Basel II and Basel III) | Basel Accords (Basel II and Basel III) | Basel Accords (Basel II and Basel III) |

Source: (BIS, 2024b, 2024a)

The Merton, (1974) modeling assesses the credit risk of a representative corporation by considering the value of its assets, liabilities, and equity. The model assumes that a company defaults if its asset value falls below its debt obligations. This model was the platform for several models that provide an estimation of credit risk, the following table provides a comparison among the main commercial models based on the Merton one.

**Table 4. Description and main features of main credit risk models**

| Model | KMV PM | CreditMetrics | PRT | CPV Macro | CreditRisk+ |
|---|---|---|---|---|---|
| Definition | Estimates probability of default (PD) based on structural approach, comparing | Quantifies credit risk using historical data, focusing on | Focuses on portfolio-level credit risk management, assessing risk | Analyzes credit risk at macroeconomic level, considering factors like GDP | Employs advanced statistical techniques to model default probability, often used |

| | firm's asset value to liabilities. | credit exposure and volatility. | across multiple assets or investments. | growth, inflation, and unemployment. | for stress testing and scenario analysis. |
|---|---|---|---|---|---|
| **Main Features** | Utilizes asset valuation and default boundary to assess PD. | Calculates Value-at-Risk (VaR) based on historical data. | Evaluates risk at the portfolio level, considering diversification. | Incorporates macroeconomic variables to assess credit risk. | Incorporates complex statistical models for robust risk assessment. |
| **Differences** | Incorporates structural approach, focusing on asset value and default boundary. | Relies on historical data and statistical analysis to measure credit risk. | Designed for managing credit risk across portfolios, not individual entities. | Focuses on systemic risk and impact of economic factors on credit markets. | Focuses on advanced modeling techniques for comprehensive risk analysis. |
| **Originator** | KMV | JP Morgan | S&P | McKinsey | Credit Suisse |
| **Year of release** | 2001 | 1997 | 2003 | 1998 | 1997 |
| **Risk type** | Default loss | Market Value | Market Value | Market Value | Default loss |
| **Credit Event** | Default | Default migration | Default migration spread | Default migration | Default |
| **Risk driver** | Asset value | Asset value (country/ industry) | Asset value (country/ industry) | Macroeconomic factors | Sector default intensities |
| **PD Correlation** | Asset value factor model | Equity value factor model | Asset value factor model | Macro-economic factor model | Default intensity model |
| **LGD Distribution** | Beta | Beta | Beta | Random | Constant |
| **PD/LGD Corr.** | No | No | Yes | No | No |
| **Calculation** | Monte Carlo simulation | Monte Carlo simulation | Monte Carlo simulation | Monte Carlo simulation | Analytical solution |

Source: (Baesens & van Gestel, 2009b)

Finally, the Basel II and III Capital Accords calculate the risk of the bank using a simplified portfolio model calibrated on the portfolio of an average bank. In addition, the Basel III Capital Accord encourages banks to measure their portfolio risk and determine their economic capital internally using portfolio models (BIS, 2024b).

## II.1.2. MARKET RISK

Market risk pertains to the possibility that the bank's financial instruments or assets invested in securities may experience losses due to volatility in market prices. Banks accumulate market risk due to their ownership of positions in financial instruments and their execution of trades on their own account. Their own capital and bank account holders are exposed to this risk. A bank's failure to effectively manage market risk can lead to substantial and immediate repercussions, such as damage to its reputation and profitability. After examining the trading instruments utilized by banks in their trading activities and the origins of market risk, the methodologies outlined in the market risk Amendment to the Basel I Accord and the Basel II Accord are included in the market risk management and measurement metrics of this framework (Skoglund Jimmy & Chen Wei, 2015).

Banking institutions can adopt both long and short positions in a variety of financial products, such as derivatives, currencies, stocks, bonds, loans, and commodities. Equity risks (representing bonds and loans), interest rate risks (representing bonds and loans), commodities risks (representing commodities), and exchange rate risks (representing currencies) are the four distinct kinds of market hazards. By implementing risk hedging strategies, financial institutions can mitigate and restrict market risk (Allen Steven, 2013).

Among the modeling approaches, the most used is Value-at-risk or VaR, which is a quantitative methodology utilized to approximate the magnitude of financial losses that are anticipated to transpire within a designated period, relying on a specific level of assurance. This procedure generates and quantifies an estimate of the quantity of risk assumed by the bank.

As part of the VaR calculation process, the present holdings of the bank portfolio must compute a range of historical return values of the portfolio performance and produce future estimations over a narrow window of time (usually one day for market risk). This graph illustrates a typical distribution of returns that may be used to determine a portfolio's potential return values. The X-axis is used to display the potential increase and decrease in returns horizontally. The sum of the curve value must equal one because it is a probability. The height of the curve corresponds to the probability that a certain gain or loss will occur at a given time at a given point. Losses are represented by negative values to the left of zero, and gains by positive values to the right (Apostolik et al., 2009).

**Graph 1. VaR modeling for market risk**

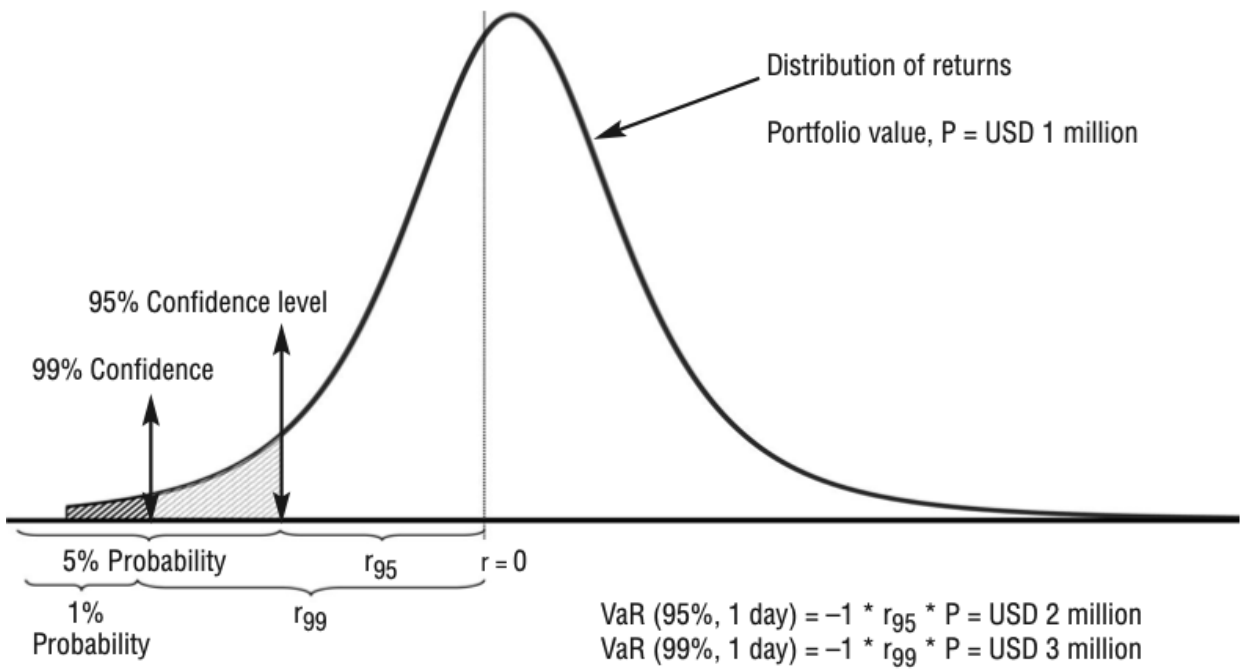

Source: Apostolik et al. (2009) "Foundations of Banking Risk." Wiley Finance.

Key assumption of market risk modeling is based on the Efficient Capital Markets Theory published by Eugene Fama (Fama, 1970a), the theory states that prices reflect all the information available in the market. According to (Fama, 1970b), the efficient capital market theory suggests that security prices at any given time fully incorporate all available information. The theory encompasses various forms of tests, such as weak-form tests that rely on historical prices, semi-strong-form tests that consider publicly available information, and strong-form tests that assume prices reflect all available information in the markets.

The theory also acts as a standard for evaluating deviations from market efficiency. In general, the efficient capital market theory focuses on whether prices accurately incorporate all available information at any given time. This theory has significant practical implications for how security prices behave. Therefore, fluctuation in prices managed by market risk management are expectations and new relevant information relevant for financial security.

<div align="center">II.1.3. LIQUIDITY RISK</div>

Liquidity risk concerns the possibility that a financial institution may suffer financial losses because of its inability to promptly purchase or sell a security held in its financial portfolio at an equitable price. The uncertainty of the bid-ask spread gives rise to this risk, specifically for securities that are thinly traded or in emerging markets during unfavorable market conditions. It is critical to model liquidity risk, as failure to do so may result in an underestimation of market risk, particularly for investment portfolios that include securities from emerging markets (Skoglund Jimmy & Chen Wei, 2015).

There are primarily two types of liquidity risks: exogenous and endogenous. The risk manager bears responsibility for endogenous liquidity risk, which stems from the sudden divestment of substantial positions that the market cannot easily absorb. Exogenous liquidity risk, on the other hand, relates to variables outside the control of the market maker. It often has significant magnitudes and affects market participants regardless of their scale (Allen Steven, 2013).

VaR methodology provides the modeling framework of the risk. According to (Van-Deventer Donarld R. et al., 2013), when evaluating liquidity risk using the VaR method, both exogenous and endogenous liquidity risk are factored into the calculation. The methodology incorporates the distribution of observed bid-ask spreads of security prices, which is readily accessible in the data, in order to quantify exogenous liquidity risk. In addition to the average spread and spread volatility, the VaR calculation accounts for the most extreme spread situations. Through the integration of these liquidity risk adjustments with conventional value-at-risk (VaR) calculations, the methodology offers a more precise evaluation of market risk as a whole, specifically with regard to substantial liquid positions.

II.1.4. OPERATIONAL RISK

According to the Basel II and III Accords, operational risk is divided into five broad categories of operational risk: 1) Internal process risk, 2) People risk, 3) Systems risk, 4) External risk, and 5) Legal risk.

**Table 3. The main risk considered as part of the operational risk**

| Type of Risk | Definition |
|---|---|
| External Risk | Risks arising from factors outside the bank's control, such as economic conditions, market volatility, and geopolitical events. |
| Internal Process Risk | Risks associated with inefficient or ineffective internal processes, procedures, or operations. |
| Legal Risk | Risks related to legal and regulatory compliance, including potential lawsuits, regulatory fines, and non-compliance penalties. |
| People Risk | Risks associated with human resources, including employee misconduct, labor shortages, and inadequate training. |
| Systems Risk | Risks related to technological infrastructure, including cybersecurity threats, system failures, and data breaches. |

Source: (Apostolik et al., 2009)

Operational risk is calculated through methodologies such as operational loss events analysis. Considering the statistical frequency analysis, the assessment of historical events is developed for the institutions assessed or through industry historical records. The analysis assumes that there is a probability distribution that fits the data assessed and there is enough information to rely on asymptotic estimation probability.

The modeling of operational risk is concentrated on two loss categories, as illustrated in Graph 2: high-frequency/low-impact events, which emerge frequently but have minimal impact or severity; and low-frequency/high-impact events, which occur infrequently but have significant consequences. The opposite extremities are typically of no concern to banks: low-frequency, low-impact events would be more expensive to manage and monitor than the losses they would cause, while high-frequency, high-impact events would indicate a bank that was fundamentally mismanaged and failed. Operational risk management should aim to ensure, as depicted in Graph 2, that high-severity operational risk events occur infrequently and those that occur frequently have an exceptionally low severity.

**Graph 2. Loss intensity and frequency of operational risk events**

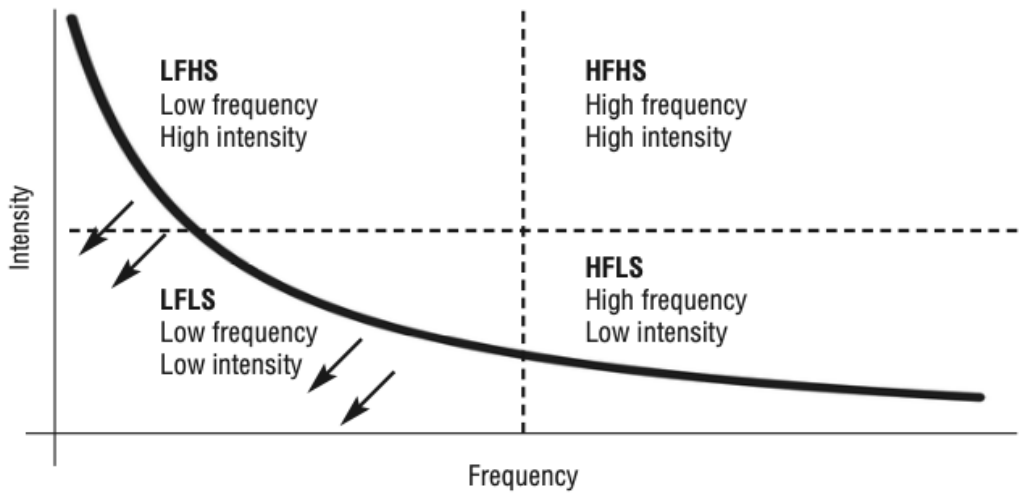

Source: (Apostolik et al., 2009)

<div align="center">

III. FINANCIAL CLIMATE RISK ESTIMATION

</div>

According to the (TCFD, 2023) and (IFRS, 2023), financial climate risk encompasses the possible adverse consequences that companies and organizations may face due to climate change and the shift toward a more environmentally responsive economy. The hazards may be classified into two primary categories: transition risks and physical risks. Transition risks arise from the significant legislative, legal, technological, and commercial changes needed to tackle the demands of climate change mitigation and adaptation.

Conversely, physical dangers arise from the direct consequences of climate change, including severe weather events (e.g., storms, floods) and gradual changes in climate patterns (e.g., increasing temperatures, anomalies in precipitation). Companies may face financial consequences due to climate-related hazards, such as physical damage to their assets, interruptions in their supply chains, alterations in water availability, and adverse effects on food security and staff safety. Furthermore, financial climate risks possess distinct attributes such as their varying impacts depending on geographical location and human activities, extended timeframes and lasting consequences, unprecedented and uncertain nature, fluctuating intensity and non-linear dynamics, as well as the requirement for temporal alignment, proportionality, and consistency in risk management procedures.

### III.1. PHYSICAL RISK

Physical risk modeling is rooted in catastrophe (CAT) modeling originally developed by and for (re)insurers (ASB, 2021). Initially, CAT modeling was used  improve the understanding of potential losses because of catastrophes. Historical CAT events showed that some of them could potentially erode drastically and even wipe out the capital base of (re)insurers.

According to (Pita et al., 2015), the conceptual framework for catastrophe modeling was initially established by Blaise Pascal during the 1660s. Pascal provides reassurance to individuals who are overly apprehensive of lightning in his work "La Logique ou L'Art de Penser." He asserts that "the fear of harm should not only be proportional to the severity of the consequence but also to the likelihood of its occurrence". This statement encapsulates the fundamental elements of contemporary catastrophe risk: the severity of harm resulting from the hazard, the likelihood of the hazard (lightning) materializing, and the degree of personal exposure (exposure) that influences the risk perception (fear of harm).

Scientific studies of natural hazards swiftly progressed during the early decades of the twentieth century (AON, 2024). However, it was not until structural engineering knowledge and the development of climatology and meteorology that an insurance company decided to adopt such innovations. (Pita et al., 2015) report that in 1954, Travelers Insurance Company initiated the establishment of the Weather Research Center, an institution to investigate the correlation between weather phenomena and various factors including agricultural losses, property damage, transportation conditions, and accident causation (Weatherwise, 1954). The Center was founded in 1955 by Thomas F. Malone, a scientist widely recognized for his ground-breaking contributions to meteorology (Byers et al., 1951). Following in Malone's footsteps, Don G. Friedman maintained a comparable association with the Center and made significant contributions to the domains of catastrophe modeling and building vulnerability estimation.

Other prominent analyses in this field include work on flood hazard published by the U.S. Water Resources Council in 1967, the Algermissen study which examined the risk of earthquakes in 1969, and the hurricane forecasts issued by the National Oceanic and Atmospheric Administration (NOAA) in 1972 (Neumann). In an effort to estimate the effects of earthquakes, cyclones, floods, and other natural catastrophes, U.S. researchers compiled hazard and loss assessments in response to these developments. Steinbrugge's compendium of losses caused by earthquakes, volcanoes, and tsunamis (1982) and Brinkmann's summary of hurricane hazards in the United States (1975) are notable compilations.

(Grossi Patricia & Kunreuther Howard, 2005) state that CAT modeling became feasible due to the convergence of two separate advancements, namely risk mapping and hazard measuring, in the late 1980s and early 1990s. Computer-based models were created by linking scientific studies on the magnitudes of natural disasters and historical events with advancements in information technology and geographic information systems (GIS). These models were used to evaluate the likelihood of severe damages and losses. The models approximated the losses caused by catastrophes by overlaying the at-risk properties with the probable sources of natural hazards in the specified geographic area. GIS has become an ideal platform for conducting hazard and loss assessments simply and cost-effectively, thanks to its ability to store and handle large amounts of geographically referenced data.

**Figure 1. Main inputs for catastrophe models**

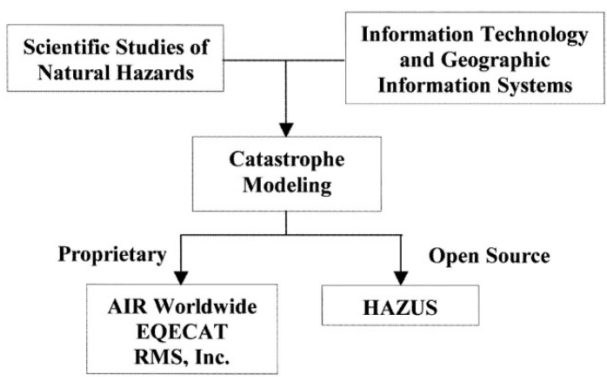

Source: (Grossi Patricia & Kunreuther Howard, 2005).

By the 1990s, the majority of major (re)insurers in Europe and the US had either established or were in the process of developing CAT models. Simultaneously, some growing modeling firms developed computer tools specifically intended to evaluate the consequences of natural hazard risk. Risk Management Solutions (RMS) was launched in 1988 at Stanford University, AIR Worldwide was established in 1987 in Boston, ERN was created in 1996 for the Latin-American market, and EQECAT was founded in 1994 in San Francisco as a subsidiary of EQE International. ABS Consulting merged with EQE International in 2001. In 2013, CoreLogic acquired EQECAT and repurposed its technology for the specific market of buyer-side modeling. This move consolidated the modeling businesses globally, with a concentration on RMS and AIR Worldwide. In 2012, one of the founders of AIR Worldwide created Karen Clark Company as the fifth modeling firm.

The introduction of physical modeling as a concept facilitated a subsequent wave of consolidation in the sector. Following the surge in financial climate risk instigated by TFCD recommendations, rating agencies, reinsurers, brokers, and intermediaries acquired modeling firms to safeguard their a operations and extend their coverage to financial climate risk, primarily in the form of physical risk. RMS was acquired by the credit rating agency Moody's in 2021. AON, an international reinsurance broker, acquired ERN in 2022. AIR Worldwide changed name to Veriskwhen acquired by Verisk Analytics in 2022. The objective of this surge of consolidation was, among other things, to reorient the sector's business scope to include financial climate risk as a top priority.

**Box 1. Weather derivatives**

According to (Considine 2000; Jewson 2004; Buckley et al. 2001),climate modeling for commercial purposes has not only been developed by (re)insurers and modeling firms but as part of the energy and agricultural market that emerged in the 1990s. It arose in a complementary way to the insurance markets. In the capital markets, specifically in the derivatives market, a first-of-its-kind contract was agreed on in the form of an option, where the conditions of a climatic variable were established, which, if it reached a pre-established threshold, triggered a payment, in a similar way to insurance. In this case, loss verification was limited to verifying that the parametric variable did or did not reach the threshold. The contracts evolved to not only be variable but to take the form of indices, known as weather derivatives (which laid the foundation for what would later be called index insurance).

Weather derivatives emerged in the context of a very warm El Nino Northern Hemisphere winter from 1997-1998 and the deregulation of the United States energy market in the mid-1990s, which led to the development of the first weather derivative contracts as an option contract.

As (Considine, 2000) notes, Aquila Energy and Consolidated Edison Co. first traded over-the-counter (OTC) weather derivatives in July 1996. Aquila Electric Power was contracted to provide Consolidated Edison with electric power during August of that year. The price of the electricity was mutually agreed upon, but the contract included a weather clause. 8 years later, weather derivatives were dominated by energy transactions, 69% of the weather market consisted of energy companies in 2004.

Because of its seasonal patterns, tendency to return to its average state, and high volatility, the weather is a distinctive "underlying asset" in the financial market.

Academics and practitioners faced a formidable challenge when attempting to model and price weather derivatives because of their unique characteristics. Besides, weather has no actual market value, so it cannot be valued in a financial market.

Seasonal patterns and the tendency for conditions to return to average can be used to predict weather to some extent. Despite this, ordinary financial markets find them to be more challenging. Meteorological variables can now be modeled using stochastic models.

By reducing their exposure to weather-related risks, climate derivatives allow corporations to reduce the fluctuations in their stock prices. The lack of correlation between weather derivatives and the stock market makes them a favorable investment.

Figure 2. Weather derivatives traded on the Chicago Mercantile Exchange

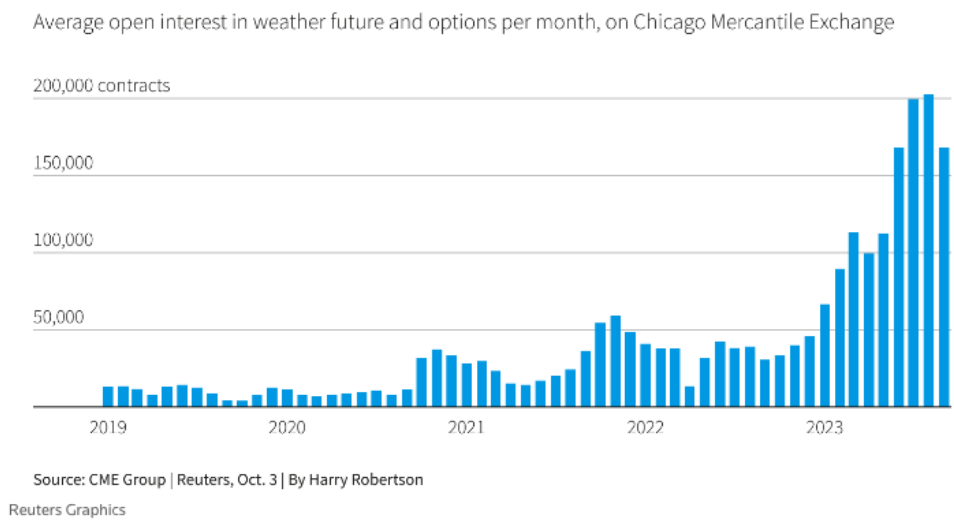

Average open interest in weather future and options per month, on Chicago Mercantile Exchange

Source: CME Group | Reuters, Oct. 3 | By Harry Robertson
Reuters Graphics

These types of models now address climate change by incorporating climate models and recent observed temperature anomalies. For example Bressan and Romagnoli (2021) Modeling climate change in weather derivatives using copula-based pricing methodologies for multivariate weather derivatives. This modeling approach is especially relevant given the impact of climate change on the pricing and trading of weather derivatives, as it introduces a further element of uncertainty in financial markets. The use of copula models can help address the challenges posed by climate change, such as heavy tail inter-dependency, asymmetric structures, hierarchical impacts, and contagious effects. Additionally, the copula model can assist in capturing the features of mean reversion and seasonality, which are important factors in climate risk hedging.

### III.1.1.  CAT MODELING APPROACH

The now classic modules of a catastrophe model are hazard, inventory, vulnerability, and loss. These components are shown in Figure 3. Initially, the model assesses the risk associated with natural hazard events. For instance, a hurricane is defined by its anticipated trajectory and maximum wind speed. The hazard may alternatively be described through the frequency of certain magnitudes or frequencies of events.

**Figure 3. Modules of a catastrophe model**

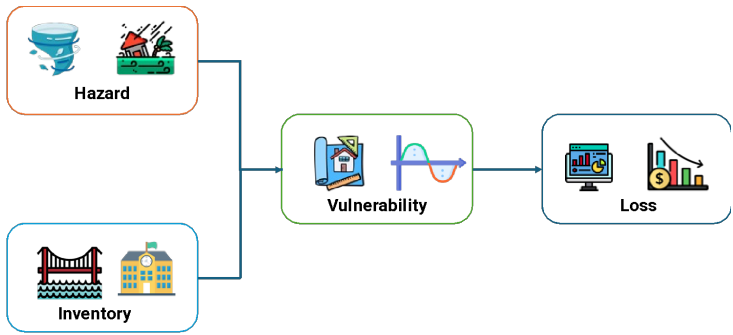

Source: based on Grissi et al. (2006).

In terms of the inventory, the model describes the assets or portfolio of properties at risk. The location of each asset is a key variable to define the level of exposure to the hazard. Geocoding the asset or proxy through certain areas (e.g., postal code) is the most used technique for location description. Additionally, as part of the data gathering for each asset, key variables should be collected that describe the vulnerability of the asset to each natural hazard assessed (e.g., construction material, number of stories, age, among others[1]).

The estimation of the vulnerability or susceptibility of the structures that are at risk of potential damage is performed by the hazard and inventory modules. The vulnerability module estimates and calibrates damage functions that set the link between a specific natural hazard on a specific physical asset. The way this link is quantified differs among various models. As an illustration, the HAZUS model classifies a structure's injury into four categories: Slight, Moderate, Extensive, or Complete. Damage curves that establish a relationship between structural damage and a particular severity parameter, such as the maximal wind speed or spectral acceleration. For each model, damage curves are generated to represent the structure, and time-related losses, according to the modeler's perspective, based on the assumption and research considered by the modeling firm. This information is classified as a company black box or business secret.

One of the core modules of CAT modeling is the vulnerability assessments. For example, hurricane vulnerability, (Pita et al., 2015) documents the intricate relationship between the main vulnerability models, which are based on historical models that have been evolving using historic data.

The plot in (Pita et al., 2015) is a study of key vulnerability models for hurricanes are an annotated timeline with the author and publication year. The methodologies are linked together using arrows to denote the interrelation between them. The perpendicular positions in the graph indicate the classification of the methodology, while the black to white colors tones represent the country of origin. The damage surveys that provided data for certain models are also illustrated in Figure 4. The timeline features prominent historical cyclones that are of particular significance. While not exhaustive, Figure 4 illustrates the most emblematic methodologies that have been devised over the past half-century.

According to (Pita et al., 2015), Figure 4 shows the influential contributions of Don G. Friedman, referenced by (Culver et al., 1975), (Clark Karen, 1986), and (Sill & Kozlowski, 1997). The studies conducted by (Sparks et al., 1994) and (Bhinderwala Shiraj, 1995) are particularly notable as they provided the data and/or assumptions that most models use. These models include the works of Walker ((Walker George, 2011)), (Huang et al., 2001), (Sill & Kozlowski, 1997), (Sciaudone J et al., 1997), (Unanwa et al., 2000), (Vickery et al., 2006) and. In Australia, the influential study was conducted by George Walker, who presented his research on wind vulnerability curves for Queensland dwellings at the Alexander Howden Reinsurance Brokers in 1994. Subsequently, (Stewart, 2003), and (Crompton & McAneney, 2008) incorporated Walker's assumptions and/or data into their own work.

The plot highlights that original vulnerability models were built exclusively from past-loss data up until the mid-1990s, as this was the only information available at the time. After the 1990s, few models were built from past-loss data only. Today many of them take into account past information for simulation, and most of them are strongly based on probabilistic simulation methods.

---

[1] When the property is insured, details on the policy's features, such as the deductible and coverage limit, are also documented.

**Figure 4. Evolution of key vulnerability models on the hurricane timeline**

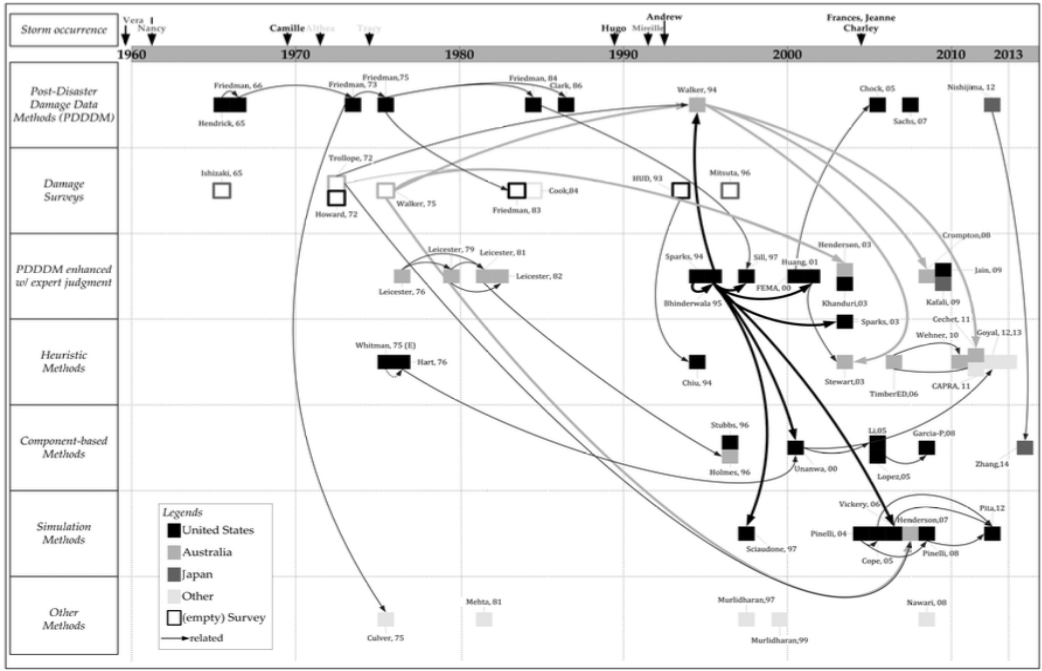

Source: (Pita et al., 2015).

An inventory loss assessment is determined by this measure of vulnerability and is estimated by vulnerability functions (Figure 5). Catastrophe models can put losses into direct or indirect categories. Direct losses include those resulting from the repair or replacement of the facility. Indirect losses include costs borne by residents forced to leave their homes and adverse impacts on companies caused by disruptions in their operations. It is possible to precisely allocate losses using exclusive models by meticulously analyzing insurance policies.

**Figure 5. Vulnerability functions and loss estimation**

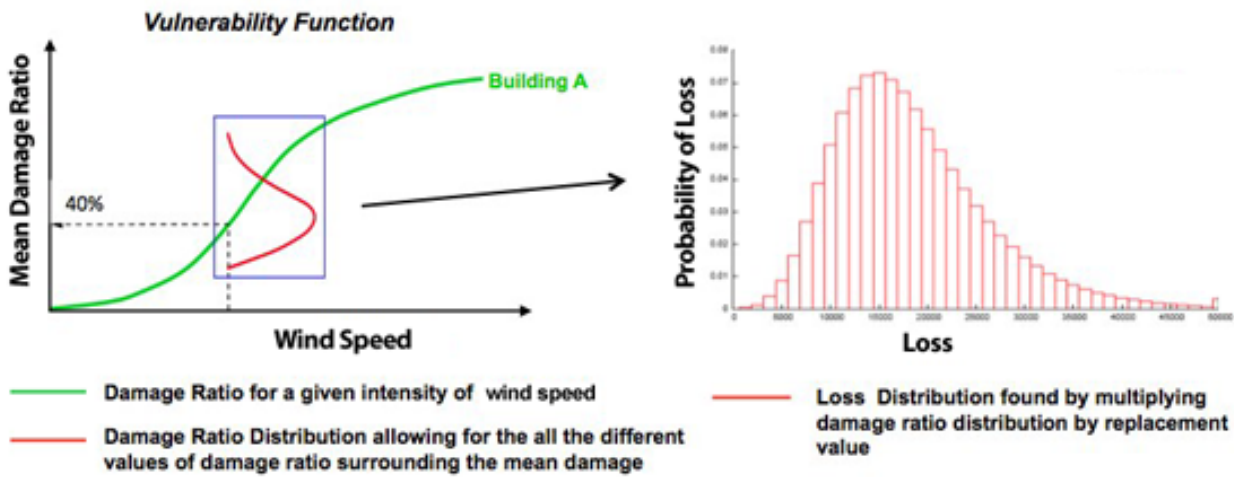

Source: Latchman (2022).

According to (CAS, 2020; Grossi Patricia & Kunreuther Howard, 2005), the final output of the loss module presents the results in terms of an exceedance curve. This curve shows, on the horizontal axis, the probability of excess of loss for a particular hazard. For each probability, there is a probabilistic value of losses. On the vertical axis (see Figure 6 left side), the zero is on the left side and tends to infinity toward the right side. On the other hand, the vertical axis starts from zero to infinity in the direction from bottom to top.

Note that on the left side of Figure 6, the curve described is ascending and asymptotic to zero, i.e., as the values of the vertical axis (probability of exceedance), the probabilistic losses tend to infinity and are capped by the exposed assets value.

Figure 6, on the left-hand side, shows an exceedance probability (EP) curve, which is the likelihood that a loss of any given size or greater will occur in a given year. In other words, An EP curve is marked to show a 1% probability of having losses of $100 million (USD) or greater each year.

Another method of expressing EP probability is the Return Period, which describes the expected likelihood of a loss of a given size occurring within a given timescale. As an example, a 50-year return period states that, on average, an event/scenario will on average repeat itself once every 50 years when repeated samples are taken.

Then, Figure 6, on the right-hand side, shows the same information as the EP Curve (left-hand side), but in terms of the return of period. To switch between these two metrics, follow these metrics: Loss Return Period = 1/(Exceedance Probability) and then, Exceedance Probability = 1/(Loss Return Period).

**Figure 6. Exceedance probabilistic curve**

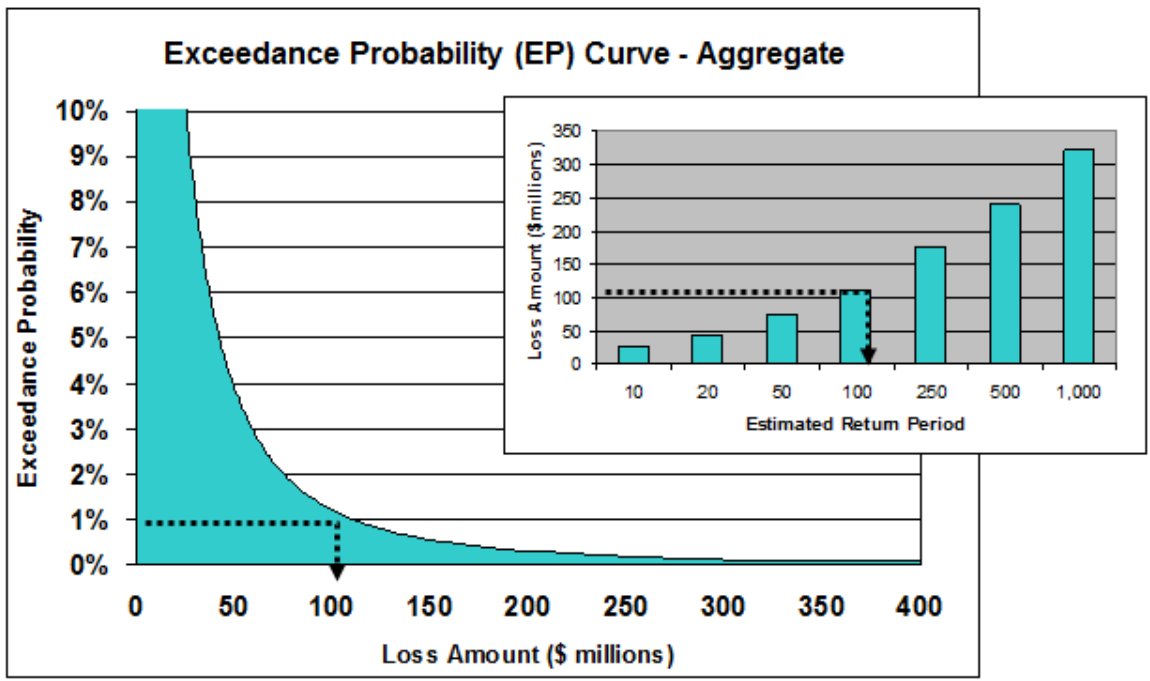

Source: (Grossi Patricia & Kunreuther Howard, 2005).

The final product of CAT modeling is utilized by (re)insurers to integrate probabilistic losses; these organizations incorporate this data into the estimation of premiums for reinsurance contracts or/and the determination of the size of catastrophe reserves in accordance with regulatory frameworks pertaining to solvency. Regarding physical risk in the context of financial climate risk, the objective of this analysis is to evaluate the effect of climate change on a company's financial performance and subsequently reevaluate the value to investors.

The 2019 Network for Greening the Financial System (NGFS) report assessed how the economic impact of climate-related physical and transition risk drivers can be transmitted to the financial system through a series of direct and indirect transmission channels.

**Figure 7. Physical Risk Drivers**

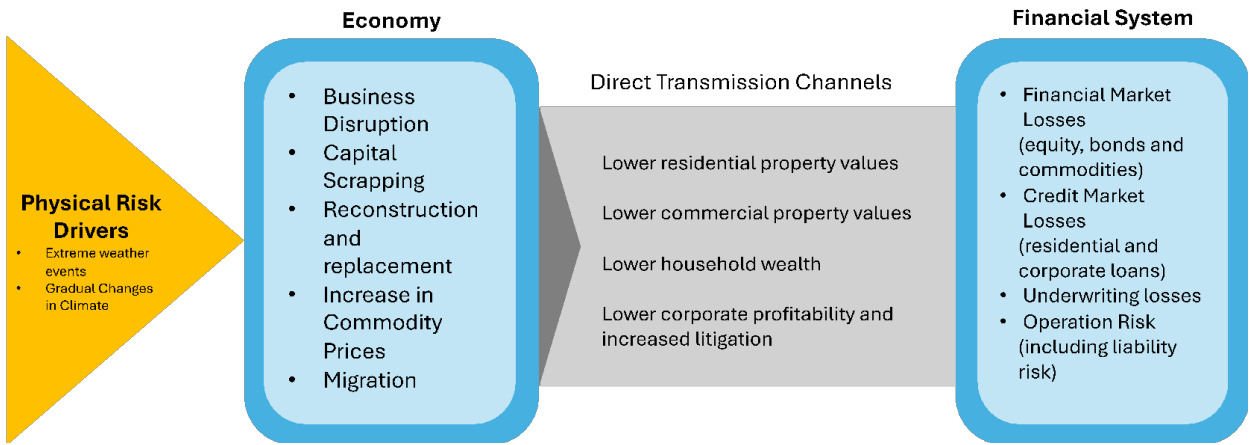

Source: (Battiston, 2022).

Financial entities could potentially be significantly influenced by financial climate risk, the Financial Stability Board (FSB) assessed in a November 2020 report. In April 2021, a report was released by the Bank for International Settlements (BIS) that is dedicated to examining the banking industry's vulnerability to physical and transition risks associated with climate change. Through comprehensive analyses of microeconomic and macroeconomic transmission channels, (Battiston, 2022) assesses the potential repercussions of climate-induced physical and transition risks on financial institutions. As summarized below (Figure 8), climate-related financial risks can be accounted for using conventional risk categories, according (DTCC, 2023) report, describes the above-mentioned risks plus liability risk, this last one implies the potential financial loss that an entity might face if it is held responsible for contributing to or failing to mitigate the impacts of climate change.

Figure 8. Physical Risk Drivers

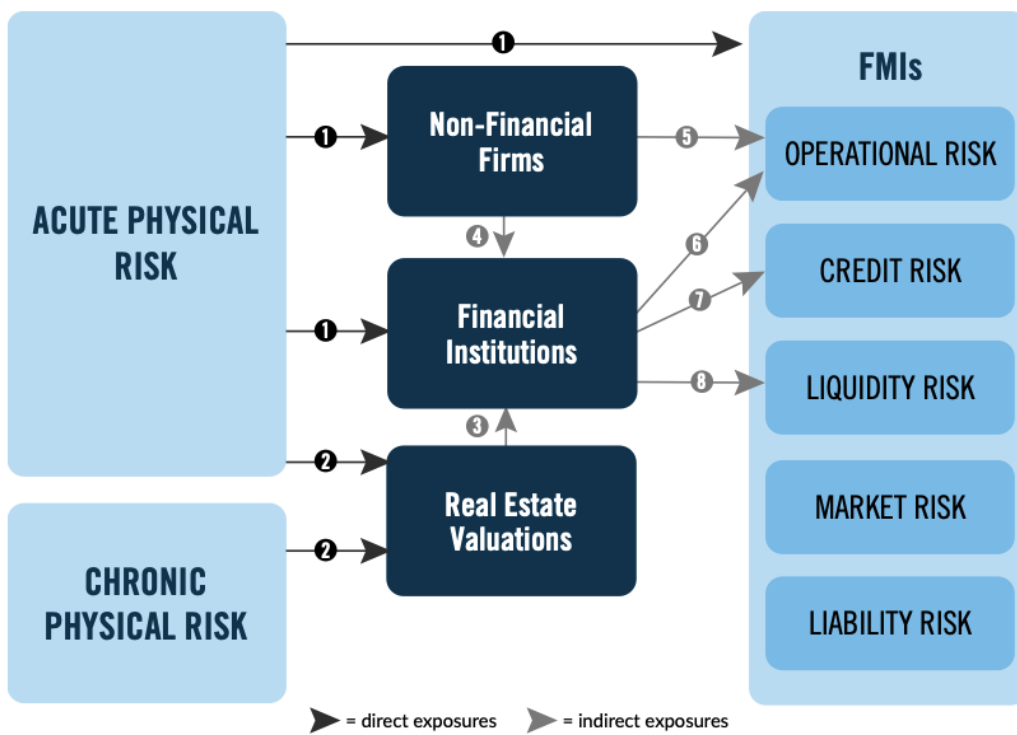

Source: (DTCC, 2023)

Physical risk estimation methods lack a standardized framework, and while these methods are based on CAT modeling, they yield varying outcomes.

Considering the scenario approach for Representative Concentration Pathways (RCP) and Shared Socio-economic Pathways (SSP), the hazard modeling should account for the estimation of physical risk as depicted in Figure 8. These scenarios are also input into the Inter-Sectoral Impact Model Intercomparison Project (ISIMIP), whose analysis serves as a tool for both hazard and vulnerability assessment.

The inventory analysis, conversely, has been extended to encompass an examination of effects that spread across the entire value chain. As a concluding step, probabilistic losses by scenario type are frequently used to represent the ultimate result.

In conclusion, it is imperative to analyze the loss outcomes concerning their influence on the organization's financial performance. Despite the existence of a few analytical frameworks, such as the Climate VaR, this final aspect of the analysis, in terms of the effect on financial performance, is still relatively uncharted territory in terms of methodological standards to be applied in general terms to any industry worldwide.

**Figure 8. Estimation of the physical risk**

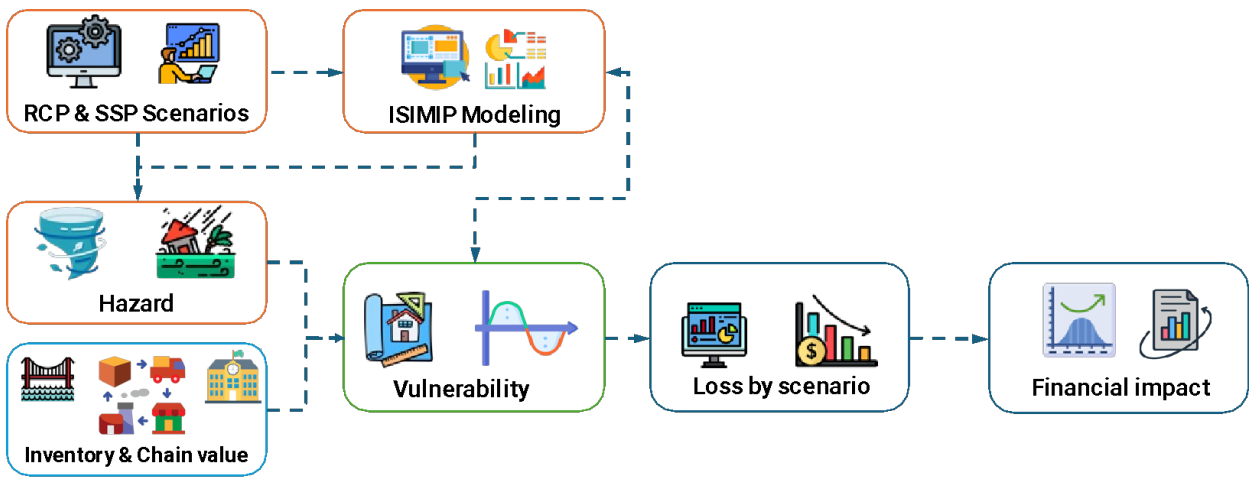

Source: author modified from (Grossi Patricia & Kunreuther Howard, 2005).

III.1.2. CRITIQUES OF THE CATASTROPHE MODELING APPROACH TO PHYSICAL RISK ESTIMATION

Recently there has been a wave of criticism about the role and relevance of catastrophe models for analyzing climate impact in the financial sector. This criticism from climate modeling companies (Jupiter Intelligence, Climate X, Riskthing.ia, among others) emphasizes that there are indeed important drawbacks or areas of opportunity. However, the research on which their criticisms are based is not yet fully understood or made public to support how to overcome these drawbacks, given the modeling is propetary and confidential, therefore only clients[2] can know those scientific foundations.

The following are the main criticisms of the use of catastrophe models to analyze climate change risk taken from Jupiter Intelligence[3] one of the main climate modelers:

- It essentially reflects the years and events that prevail in the present day. As a result, evaluating future events that have not been observed in the past is impossible. In the future, events might also aggregate and compound in a manner distinct from its current state.
- The analysis fails to account for more extensive climate change. As a result, severity and frequency are typically underestimated.
- Limited perils and regions. Cat models, because the business model is focused on main insured markets geographically, miss other regions relevant for a comprehensive analysis of global climate.

---

[2] Among those clients are, the US government, the largest bank and financial players, as well as oil producers.
[3] https://www.jupiterintel.com/

- Slow to update. Catastrophe model updates typically dwell more on the impacts of climate rather than on the assets being analyzed.

Finally, another major criticism (Turner, 2023) is that catastrophe models assume stationarity in the time series, which is questionable as a working assumption precisely because of recent changes in climate anomalies.

### III.1.3.   LITERATURE REVIEW OF PHYSICAL RISK

The existing body of literature concerning physical risk can be broadly categorized into two groups: those that provide explicit (*E*) estimates of physical risk as per the TFCD definition, and those that implicitly (*I*) state that their estimates are physical risk estimates. This second group may make an indirect reference to the definition of TFCD, it does not explicitly state that it represents a physical risk estimation (see Table 4).

Physical risk estimation presents an enormous challenge for the scientific community. Following international practices that are fundamentally derived from catastrophe modeling, it requires asset exposure databases, which are built with a specific profile considering physical attributes and location, among other factors. Obtaining a substantial portion of this information is frequently challenging due to its proprietary or confidential nature. The design and calibration of damage and vulnerability functions to model the asset portfolio, is an additional obstacle that is typically extremely onerous due to the complexity of developing, compiling, and parameterizing under the constraints of the database, considering the calibration adjustments that are made in consideration of the specific characteristics.

For this reason, the emerging body of literature consists primarily of research conducted by the financial authorities of the countries that have generated these estimates, due to their ability to access such data. Conversely, there are corporate reports that do not achieve the extensive dissemination via academic journals. In contrast, academic research, except for a few instances, has limited access to such data or employs alternative methodologies, which we classify as implicit risk estimation.

**Table 4. Physical risk: Literature review (E = explicit and I = implicit estimates)**

| Reference | Description | Type |
|---|---|---|
| (Fiedler et al., 2021) | The document discusses the assessment of financial risk associated with the physical impacts of climate change, particularly changes in climate extremes. It emphasizes the need for knowledge of climate change across multiple spatial and temporal scales to assess these risks. The use of information from global general circulation models (GCMs) and regional climate models (RCMs) is highlighted as a key input for examining climate risks. The document also mentions the challenges in assessing physical risk, such as the difficulty in predicting extreme events and the limitations of current climate models in providing detailed and reliable information at smaller spatial and temporal scales. | I |
| (Pinchot et al., 2021) | The analysis focuses on the coverage of physical climate hazards, qualitative guidance, identification of physical climate hazards, and specific metrics provided by the disclosure initiatives. It also highlights the gaps in the guidance, such as the absence of a shared robust understanding and approach to identifying and assessing physical climate risks. The document proposes several actions to address these gaps, including the development of an open-source, science-based framework for assessing physical climate risks and the need for more accessible, practical, and open-source scientific datasets on climate-related hazards. | I |
| (Caloia & Jansen, 2021) | The document highlights the importance of quantifying the financial implications of climate-change physical risk in a structured manner and provides the first quantifications using the Netherlands as a case study. The document assesses physical risk by using a stress test framework and geocoded data to map flood areas to the level of postal codes at the 4-digit level. It starts with flood maps provided by the Dutch government, which are based on the 2007 E.U. Floods Directive. The maps developed assess flood risk, map the flood extent, and provide information to the public on the results. | I |
| (Calabrese et al., 2022) | The document assesses the physical risk by analyzing the impact of climate-induced changes on default risk in the context of mortgage lending in Florida. It considers various environmental factors such as sea-level rise, cyclonic intensity, precipitation patterns, and river discharge to forecast changes in flood risk over the next 30 years. The analysis focuses on the RCP 4.5 climate change scenario, which estimates the | E |
| | share of properties at risk of flooding under current climate conditions and projects the impact of climate change on flood risk exposure. The study also incorporates extreme weather events such as hurricanes and heavy rains to assess their impact on default probability. Additionally, the document uses a spatial additive survival approach to incorporate the effects of extreme weather events in the analysis of credit risk. | |
| (Bikakis, 2020) | The paper assesses physical risks by focusing on the impact of flooding on the UK's financial stability. It uses a scenario-based model to estimate flood risk, using flood emulation scenarios per region developed by Sayers (2015) under 2°C and 4°C scenarios in 2080. The direct effects of flood risk on financial stability are also discussed, including physical damages to properties, reduction in property value, increase in Loan-To-Value ratio (LTV), and increase in mortgage default rates. The study also considers the potential indirect effects of flooding, such as the reduction of neighboring property values, declining performance of financial institutions with no exposure, and reduction of bond ratings. | E |
| (Banque of France, 2020) | The document assesses the physical risk associated with climate change, in the context of the financial sector in France. It highlights the challenges faced by financial institutions in identifying and quantifying the impact of physical risks on their portfolios, such as immovable properties and corporate assets. The assessment is based on the IPCC "RCP 8.5" scenario, which projects a significant increase in the frequency and cost of extreme weather events due to climate change. The document also emphasizes the need for improved data collection, modeling, and methodologies to better account for physical risks at the sectoral or company level. Additionally, it underscores the importance of integrating climate risks into financial risk assessment processes and the need for collective efforts to address these challenges. | E |
| (Bank of England, 2019) | The document assesses physical variables by specifying the changes in the frequency and severity of weather events, both in terms of mean outcomes and outcomes in the tail of the distribution. It incorporates both the acute and chronic impacts of physical risks and focuses on variables that directly impact banks' and insurers' assets, as well as insurers' broader underwriting portfolios. The scenarios are based on external research that specifies these variables with a high degree of geographic granularity and at the level of the individual peril to reflect their regional impacts. The document also discusses the calibration of physical and transition risk variables, the impact of transition risks on household sector exposures, and the assessment of the vulnerability of individual counterparties to physical risks. Additionally, it outlines the approach to specifying and calibrating physical and transition risk variables, and the expectations for participants to model the impact of the scenarios on their assets at a granular level. | E |
| (Mccarthy et al., 2024) | The document assesses physical climate risk in Australian residential mortgage-backed securities (RMBS) using two risk metrics. It examines how RMBS with higher levels of physical climate risk are typically issued by small regional banks and credit unions, rather than large banks or non-banks. The analysis also investigates the relationship between climate risk exposures and credit enhancement from the subordination of junior notes. It suggests that AAA-rated notes in RMBS with higher climate risk do not benefit from additional subordination of junior notes, indicating that securitization markets may not fully incorporate climate risk exposures into their assessments of RMBS. Additionally, the document estimates the impact of increasing physical climate risk on housing prices for securitized mortgages, finding that the impact is estimated to be small. The analysis is a first attempt at quantifying climate risk present in Australian RMBS and is part of ongoing work at the RBA to assess the effect of climate change on the financial system. | E |
| (Committe Expert Group on Climate Change & Group on Securities | The document outlines the methodology for constructing physical risk indicators, which are used to assess the exposure of euro-area financial institutions to physical risks stemming from natural disasters and chronic phenomena induced by climate change. The assessment involves integrating climate data with financial datasets to develop indicators such as risk scores and expected loss (EL) indicators. These indicators are calculated for companies that are counterparties of financial institutions and aggregated to the sector and country level. The assessment also incorporates risk mitigation strategies such as collateral and insurance to reflect possible reductions in the financial consequences of exposures to physical risks. The indicators are compiled using a bottom-up approach, starting from the registered address of the company | E |
| Statistics, 2024) | concerned and linking climate data to the exposures of the company's physical assets and financial obligations. The document also highlights the limitations and challenges in assessing physical risk, such as data gaps, the need for improvements in climate models and data sources, and the difficulty in capturing the secondary effects of natural disasters. Overall, the assessment of physical risk indicators in the document emphasizes the importance of understanding and mitigating the financial system's exposure to physical risks associated with climate change. | |
| (of Japan, 2022) | Physical risks are assessed in the document through a comprehensive analysis of the impacts of floods on the real economy, land prices, and financial conditions of Japan's financial institutions (FIs). The document discusses the direct and indirect effects of flood damage, including the decline in labor productivity and the adverse impact on real GDP. It also presents estimates of physical damage to real GDP stemming from various physical risks, such as floods, droughts, and wildfires, using climate change scenarios. The assessment also considers the uncertainty surrounding the extent and pace of the increase in physical risks, which depends on factors such as the pace of transition to a decarbonized economy and the interaction between global average temperature and the frequency and scale of disasters. Additionally, the document highlights the importance of considering the future impacts of possible climate changes when assessing physical flood risks from a long-term perspective. | E |

From the point of view of macro-financial authorities, physical risk has attracted strong attention to the housing sector, (Contat Carrie Hopkins Luis Mejia Matthew Suandi et al., 2023) conduct a literature review on the impact of physical risks, such as natural disasters and chronic climate-related changes, on the housing sector. It examines the effects of physical risks on housing prices, mortgage performance, migration patterns, and insurance uptake.

The evidence of physical risk in securities pricing is a significant concern in the context of climate change. The literature highlights the impact of natural disasters and chronic climate-related changes on real estate markets, which can have direct implications for securities pricing.

Studies have shown that physical risks, such as flooding and wildfires, can lead to declines in property values and affect the pricing of securities tied to real estate assets. For example, research has found that natural disasters can lead to price decreases in housing markets, particularly in areas directly affected by these events. Additionally, the literature emphasizes the need for more comprehensive data on physical risks to improve predictions and policy responses in the housing and mortgage markets, which can ultimately impact securities pricing.

Furthermore, the potential consequences of overvaluation in housing markets due to unpriced climate risk have been highlighted, indicating the need for a deeper understanding of how physical risks can influence securities pricing in the context of climate change.

Finally, according to a literature review by (De Bandt et al., 2023), the work assesses if securities pricing reflects to some degree physical risk, the findings are listed as follows:

- There is evidence that physical risks are being priced in certain markets, such as credit and equity markets, but the evidence is preliminary and sometimes mixed.
- In credit markets, investors seem to pay a premium for corporate bonds that tend to perform better when bad climate news arrives.
- Some signs of physical climate risks being priced in sovereign debt markets have been observed, for example, extreme weather conditions causing borrowing conditions to deteriorate for sovereigns in the Caribbean.
- In equity markets, the elasticity of equity prices to temperature risks across global markets appears to be negative and increasing in magnitude over time along with the temperature rise.
- Heat stress has been robustly priced in municipal and corporate debt, and equity markets since 2013, but there is no evidence of pricing for other physical risks.
- There is microeconomic evidence for the pricing of physical risks in housing markets, with some studies finding that inundation risks are priced into residential real estate valuations, while others find no effect of inundation risk being priced into residential real estate valuations.

Overall, the evidence suggests that while there are indications of physical risks being priced in certain securities markets, the findings are not consistent across all markets and the evidence is still preliminary.

III.2. Transition risk

The precedent for transition risk estimations are studies that examine the economic repercussions of emissions but there is not a standard methodology (e.g., Nordhaus and Stern). These studies have reported their findings regarding the economic burden imposed by greenhouse gas emissions, expressed as the cost per ton of greenhouse gas emissions. Despite this, a more detailed examination is required according to the most widely accepted definition of transition risk (TCFD, 2017). As a result of the guidance provided by the Financial Stability Board and TCFD in 2017, the analysis reached a turning point.

Based on this TFCD definition, to assess transition risk, four main concepts are used, as illustrated in Figure 9: policy and legal, technology, market, and reputation. Strategic planning and risk management focus on analyzing the subjects of analysis. An income statement, a cash flow statement, and a balance sheet are all about a company's finances. In addition to assets, liabilities, capital, and alternative financing sources, it examines revenue and expenditures. Consequently, prior 2017 research conducted at the macro level did not exhibit the level of dissection mandated by the definition.

**Figure 9. Financial climate risk breakdown according to TFCD**

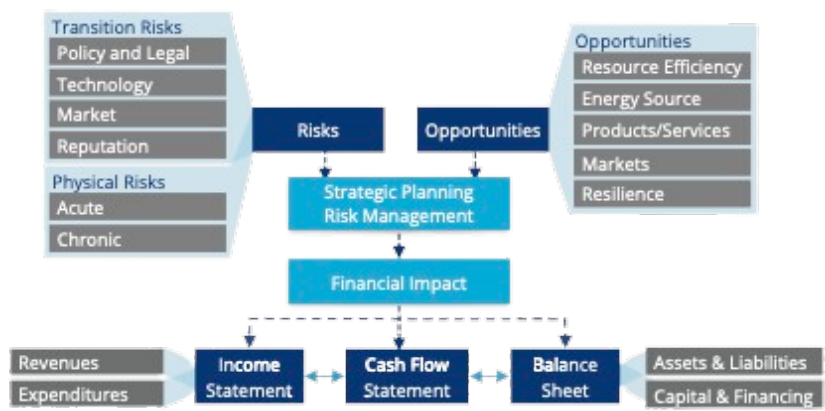

Source: (TCFD, 2017).

The authors of this new generation of models that include transition risk are generally linked to studies developed by financial authorities, including central banks. Despite the commonality of general equilibrium models, the literature exhibits a wide range of approaches, examined more closely below.

To get a first approximation of these models, (Dunz et al., 2021) and provide an example of these kinds of models, in which six sectors of an economy are modeled, households (H), government (Gov), banking (Bk), consumer goods producers (F), brown (linked to high emitters of greenhouse gasses) capital goods producers (B), and green capital goods producers (G). An interconnected network of balance sheets is used to represent sectors, in which accounting identities persist regardless of the behavioral rules. An economy with circular flows is shaped by the interactions among sector assets and liabilities via capital and current account flows.

Income for households is derived from wages and dividends from the firm sector, as well as interest on deposits and dividends from the banking sector. Taxation contributes to the government's fiscal revenues from firms, banking, and household sectors. As a result, fiscal revenues can be used by the government to cover current expenditures and public investments (such as infrastructure and welfare), as well as green and brown capital investments.

To refinance its operations, the government issues sovereign bonds, which are purchased by the banking sector. Two capital goods producers (green and brown) and one consumption goods producer make up the firm sector. Depending on the intensity of emissions of their production, capital goods producers can be classified as either brown or green. The intensity of greenhouse gas emissions is lower for green capital goods producers. Consumer goods producers can choose between green and brown capital goods.

This model, in particular, provides a contribution to the literature on climate-financial risks by explicitly modeling banking sector climate sentiments, i.e., their anticipated effect of future climate policies on green and brown firms. Therefore, according to (Dunz et al., 2021), to embed the characteristics of climate transition risks, such as deep uncertainty, nonlinearity, path dependency, and endogeneity[4]

---

[4] Endogeneity refers to the condition where variables within the model are determined by the interactions and relationships among the other variables in the same model.

(Battiston and Monasterolo, 2019a), a forward-looking approach to firm's risk assessment is required for the analysis of banking sector climate sentiments.

**Figure 9. Sample of the model framework in the context of transition risk**

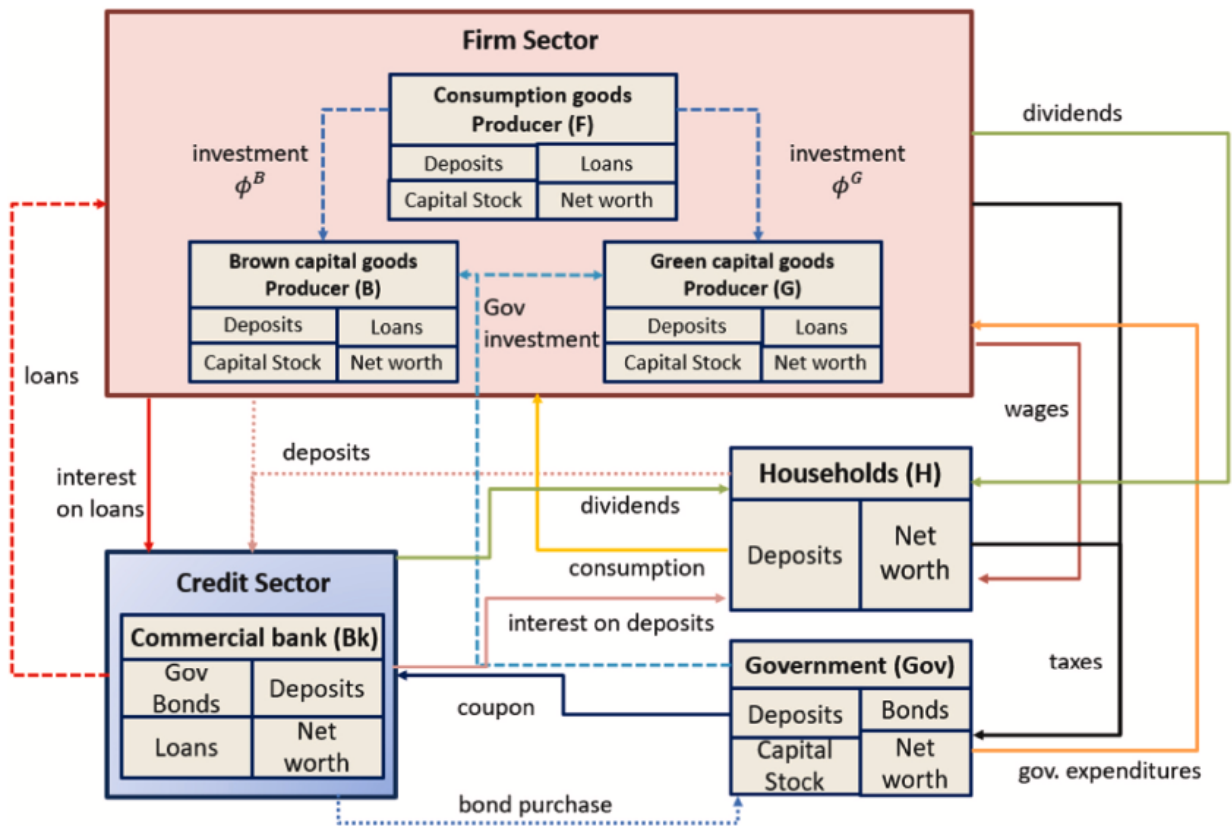

Source: Dunz et al. (2021).

However, after the achievements of new model generations that make distinctions between green and grown financial assets, the studies are approaching a deeper level of granularity, particularly in terms of the sectors for the firms assessed. This is why the use of NACE (Nomenclature des Activités Économiques dans la Communauté Européenne) has become popular among researchers, which is a European industry standard classification system.

For instance, (Battiston, 2022) propose a mapping from NACE codes of economic activities to Climate Policy Relevant Sectors (CPRS) and Integrated Assessment Model (IAM) variables (Figure 10). This mapping aims to support climate risk analysis of financial portfolios using Network for Greening the Financial System (NGFS) scenarios. The CPRS enables the grouping of NACE codes into categories of climate transition risk, allowing for the identification of the most relevant Integrated Assessment Models (IAM) variables for use in climate scenarios. This mapping provides a science-based, transparent, and operational tool to support practitioners, financial supervisors, investors, and academics in climate transition risk disclosure and assessment. The document emphasizes the importance of considering the specific technologies used in economic activities for assessing climate transition risk, as well as the relevance of CPRS in addressing the challenge of identifying potential carbon-stranded assets.

Figure 10. Mapping economic activities into AIM variables via Climate Policy Relevant Sectors

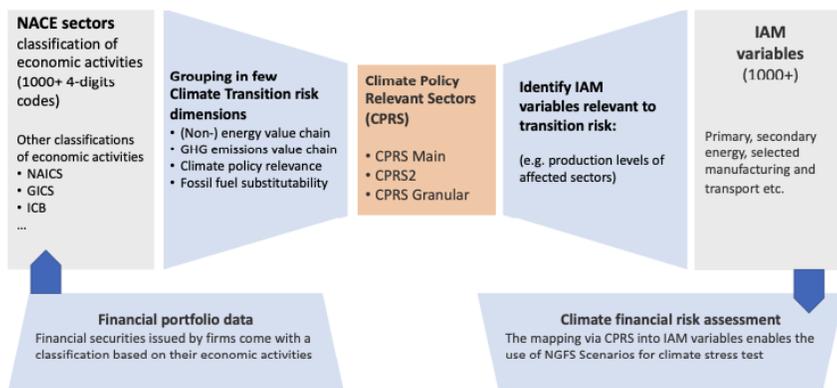

Source: (Battiston, 2022).

Combining this level of granularity in the context of a general equilibrium model is the typical approach of a central bank for assessing system transition risk for a given economy.

### III.2.1.   AN OVERVIEW OF 'TRANSITION RISK' MODELING BEFORE 2017

Between 1992 and 2017 there is no explicit definition of transition risk, in the sense that TFCD does. During this period the transition risk was implicit and was generally calculated globally for the whole economy and not per economic agent. Prior to the widespread adoption of the TCFD recommendations, the economic scientific community had been concerned with the effects of climate change emissions on economic dynamics and the assessment of consequences of public policy on specific emissions control policies since the 1990s. In particular, assessing the cost of maintaining the baseline scenario (business as usual) versus examining the repercussions on the economy shifting to lower emission patterns caused by public policies and/or market trends (Nordhaus, 1993).

In 2016, the TCFD recommendations popularized the concept of transition risk, as the potential costs to society of evolving to a low-carbon economy to mitigate climate change. However, this concept had beenassessed in 1992 in parallel with the formalization of climate change as a global threat at the UN Earth Summit in the same year.

From 1992 to 2016, four innovative models were introduced. The first of these was the "Dynamic Integrated Climate-Economy model (DICE)" in 1992 (Nordhaus, 1993). The DICE model, which is classified as a neoclassical integrated assessment model, incorporates the carbon cycle, climate science, and estimated impacts. This integration enables the evaluation of the costs and benefits of implementing measures to mitigate climate change, which are subjectively estimated.

In 2006 the Economic Projection and Policy Analysis (EPPA) model was released (Paltsev et al., 2005), a comprehensive tool used to analyze the economy and various policy scenarios. It is often employed to assess the economic impacts of policy changes, such as tax reforms, environmental regulations, or trade agreements. The EPPA is a computable general equilibrium (CGE) model, meaning it considers interactions between different sectors of the economy, households, and governments.

In 2006, The Stern Review on the Economics of Climate Change was published, based on several economic models to assess the impacts of climate change and the cost-effectiveness of different mitigation and adaptation strategies (Stern, 2008). One of the key models used in the Stern Review was the PAGE (Policy Analysis of the Greenhouse Effect) model.

In 2010 the IMACLIM-R model was released, which is a computable general equilibrium (CGE) model used for assessing the economic impacts of climate change policies, particularly focusing on greenhouse gas emissions reduction policies (Sassi et al., 2010).

In 2016, there was a notable surge in scholarly publications dedicated to transition risk, specifically defined in accordance with the TFCD guidelines. This surge prompted the initiation of a formal investigation into the overarching concept of transition risk. A substantial body of literature has surfaced that addresses various facets of transition risk. Broadly speaking, the body of literature has primarily concentrated on comprehensive analyses of the economy, specific sectors, or individual companies. A limited number of instances present a methodology that enables the examination of specific corporations.

This level of analysis may be attributable to the analysis's level of model resolution; despite being extremely comprehensive, this first set of studies hardly enables us to comprehend what transition risk is and how its roles at firm level individually. Consequently, there are few instances in which methodologies pertinent to specific corporations on a global scale can be derived.

III.2.2.    AN OVERVIEW OF TRANSITION RISK MODELING AFTER 2017

In general terms, 2017 was the tipping point for a new wave of literature on financial climate risk, specifically in terms of transition risk as defined in the TCFD. Considering (Le Guenedal, 2022) classification in his literature review, the following list breaks down the literature into two groups (before and after 2017), and then, the literature is split into terms of 8 classifications (Table 5).

Table 5. Types of approaches in transition risk literature

| # | Classification | | Description | Reference |
|---|---|---|---|---|
| 1 | Bottom-up approaches | | This subset of the literature studies at the firm level the transition risk, assessing methodological approaches on credit risk assessment, and studying the potential impact of that assessment adding transition risk. | (Howard A & Patrascu O, 2017; Nguyen et al., 2020) |
| 2 | Stress-testing | | This is one of the more prolific pieces of literature, dominated by financial institutions, and researchers focused on financial stability at the national or global level. The cascade or transmission along economic sectors is the main research topic. This field of study coincides with the main aim of financial regulators to assess any threat of financial vulnerability, in this case by climate change. | |
| | 2.1 | Financial climate stress-test | As a result of the global trend among financial authorities to evaluate the vulnerability of financial systems to climate change, the banking industry has an urgent need for stress-test assessments. This has led to the proliferation of this field of study. As part of a new generation of models, cascading effects have been recently incorporated into the design of climate stress tests. | (Allen et al., 2020; Alogoskoufis et al., 2021; Battiston et al., 2017; Dunz et al., 2021; Gourdel et al., 2021; Grippa et al., 2020; Naqvi & Monasterolo, 2021; Roncoroni et al., 2021) |
| | 2.2 | Stress-testing frameworks considering physical supply-chain interdependencies | Although the literature in this discipline is state-of-the-art, there are still numerous obstacles to overcome. This literature not only examines the interdependence among economic sectors, which is characteristic of transition risk, via the transmission channels that are commonly assessed but also investigates the interconnections between physical risk and economic sectors, as well as the interconnections with the rest of the financial industry. | |
| 3 | Direct Exposure | | This literature studies the exposure of corporations, mainly financial institutions, in their investments to fossil firms and their commodities. Financial authorities are strongly incentivized to assess this exposure in order to identify how financial entities regulated are part of the chain value of the fossil fuel industry. | (Bank of England, 2019; Batten et al., 2016; EIOPA, 2017; Giuzio et al., 2019; Weyzig et al., 2014) |
| 4 | Direct valuations | | This literature assesses the direct impact in the valuation of those securities directly exposed to transition risk, for example, those securities (equity and bonds) issued by fossil fuel firms. | (Weyzig et al., 2014) |
| 5 | Integrated assessment models (IAM) | | At the core of these models consider cost-benefit and techno-economic valuation methodologies. They enable policymakers to devise a trajectory for the social cost of carbon that optimizes long-term welfare by weighing the expenses of current mitigation against those of future damage. These models are attractive for policymakers assessing the economy on a global basis or by country and regionally. | (Nordhaus, 1993; Paltsev et al., 2005; Sassi et al., 2010; Stern, 2008) |
| 6 | Market measures | | The evaluation of asset prices as recipients of all relevant and available information during the pricing process establishes a study field that compares the transition risk impact and the degree of efficiency, i.e., the precision with which asset prices reflect the information available of transition risk. | (Apergis et al., 2022; Bennani Leila et al., 2018; Drei et al., 2019; Duan et al., 2021; Eliwa et al., 2021; Green Bonds et al., 2018; Hoepner et al., 2023; Raimo et al., |

| | 6.1 | integration of ESG in securities pricing | Within this particular domain, research is concentrated on the intersection with the Environmental, Sustainability, and Governance (ESG) literature. This literature, in turn, establishes metrics specific to ESG and evaluates security portfolios, assessing various ESG policies implemented by securities issuers and firms, while considering market metrics of portfolio value. | 2021; Riordan et al., 2020; Slimane et al., 2019; Zerbib, 2019) |
|---|---|---|---|---|
| 7 | Top-downs approaches | | For example, in a study on the financial risks associated with the transition to a low-carbon economy in the Netherlands. It includes stress tests and scenario analysis to assess the potential impacts on financial institutions and the broader economy. The study uses macroeconomic simulations and industry-level data to evaluate the vulnerabilities and exposures of financial institutions. It also discusses the limitations of the models used and the need for improved data quality. The scenarios are designed to be globally relevant and plausible in the short term, aiming to provide immediate relevance to decision-makers and stakeholders. The study aims to contribute to the understanding of energy transition risks and their implications for financial stability. | (Vermeulen et al., 2018, 2019) |
| 8 | Transmission channel | | This is one of the leading topics in the literature, based on the identification and assessment of transmission channels of the transition risk into the economic and financial performance of a given firm and at an aggregate level for a sector or an economy. | |
| | 8.1 | Impact in the cash flows | In this approach, the cash flows are assessed, considering that could be potentially affected in two distinct ways by the transition risk: either revenues decline due to decreased demand for carbon-intensive products and services, or demand increases for transition-friendly products and services emerge as a consequence of the transition. Transition policies may also have an impact on operating expenses, in addition to the direct emission costs and indirect emissions costs (carbon price) that businesses will incur from the supply chain. | (Colas J et al., 2018; Monnin, 2018; Thom J & Ralite S, 2019) |
| | 8.2 | balance sheet (stranded assets) | This literature assesses the impact at the balance sheet level, identifying and estimating the impact on the balance, for instance, losses in reserves because of stranded assets, and profit accumulation because of a slowdown in competitiveness. | (Caldecott et al., 2013, 2016; Caldecott & Smith School of Enterprise and the Environment, 2016) |
| | 8.3 | Stock market response | As a key topic in the literature, considering the abundant amount of data in financial markets in terms of daily prices on securities (mostly equity and debt), this segment of the literature studies the impact of transition risk on the pricing of these securities. | (Andersson et al., 2016; Bolton & Kacperczyk, 2023; Engle et al., 2020; Faccini et al., 2023; Gurvich & Creamer, 2022; Harris, 2015; Roncalli et al., 2021) |

Figure 9 illustrates the timeline of transition risk modeling after and before the TFCD recommendations. This literature has flourished after the TFCD recommendations, in part because of the concern of global financial authorities and some years later, the support of many governments worldwide which are in the process of making this kind of analysis mandatory.

**Figure 9. Timeline of transition risk literature development**

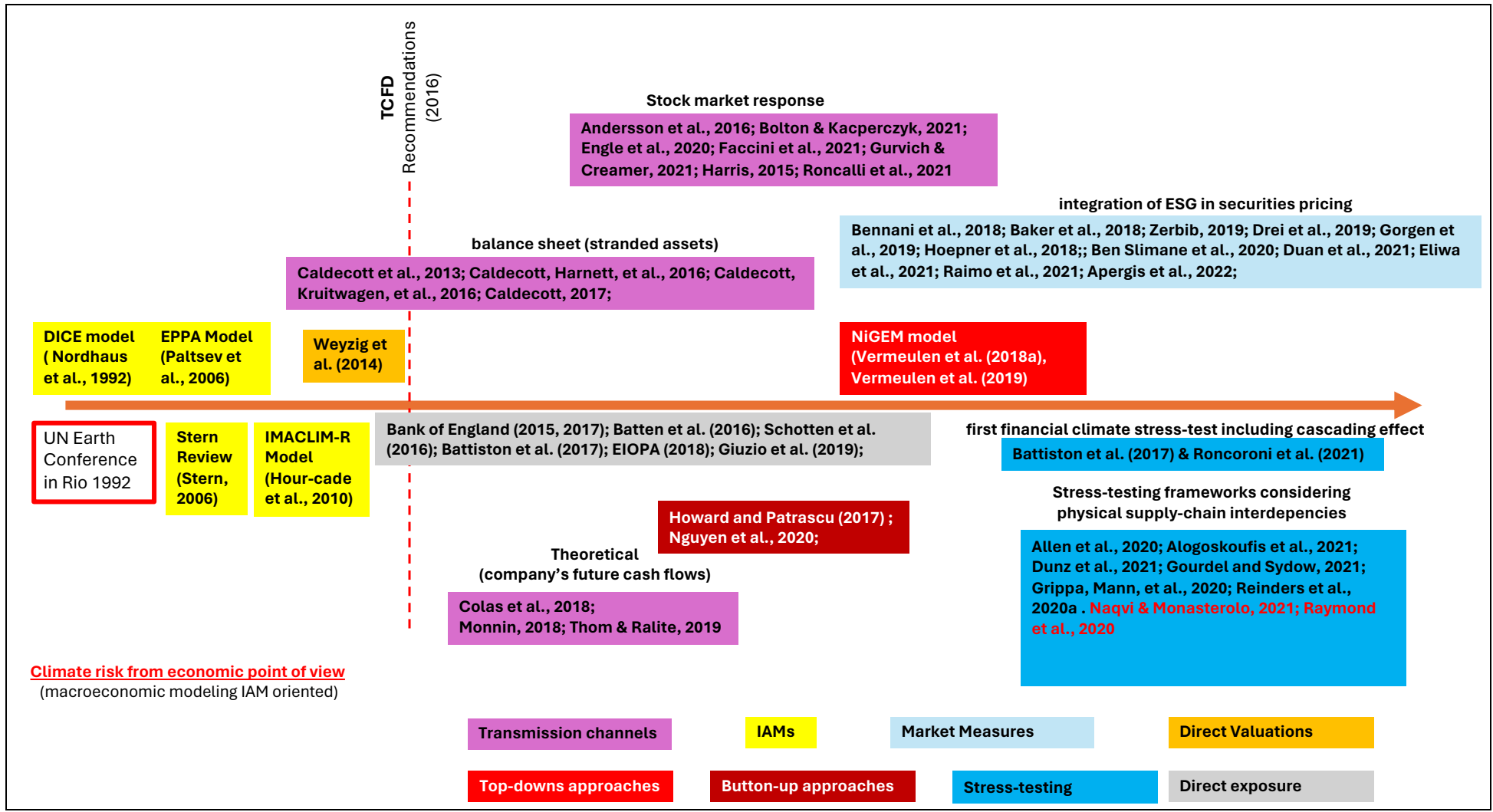

**IV.**   Conclusions

Since its proper definition in 2017, the literature on financial climate risk has experienced rapid growth in terms of transition risk as well as physical risk.

Physical risk is an example of an area where granular analysis has achieved a high level of precision, and an uncertainty assumed more transparent in the analysis. This is supported largely by catastrophe modeling studies. Notwithstanding facing numerous criticisms, the prospective agenda implies that the numerous obstacles that regulators, policymakers, and market participants in the various sectors of the economy still face might be progressively overcome.

In both risks there are significant advances in the literature, but additional challenges remain, such as the need for detailed databases of exposed physical assets and climatic hazard modeling. Despite the notable achievements made by the scientific community, challenges remain, such as designing damage and vulnerability functions based on the diversity and individual features of physical assets, thus reducing the basic risks associated with analyzing heterogeneous databases at the national or at global scale.

Physical risk studies often rely upon existing climate change scenarios, whether IPCC (RCP and SSP) or NGFS. However, this approach does not take into account probability in these scenarios. As a result, the scenario analysis lacks the probability of occurrence as a key factor in the analysis for estimating financial climate risk.

Transition risk, conversely, has transformed since IAM models centered around the macro estimation of the cost of carbon for the economy incorporated into climate scenarios and, more recently, were defined in the context of economic equilibrium models encompassing multiple sectors. Within this framework, the resolutions of IAM models have been enhanced through the incorporation of sector-specific emission assessment, taking into account NACE codes and operating similarly to the North American Industry Classification System (NAICS) and Standard Industry Classification (SIC).

The literature on physical and transition risk is converging to address compounding (and cascading) risk, a novel method gaining pace. These studies aim to capture the non-linear impacts of compound climatic shocks, which, when considered individually, often leading to an underestimation of dangers. Further study and improvement are necessary for other important models used to assess compound hazards, such as climate and disaster risk models.

Furthermore, questions remain in the literature on how to properly interpret the estimates of physical and transition risk in terms of the existing risk management methodologies within banks, particularly in small banks. This represents a key challenge, not only in the context of measurement itself; the last mile, in the long process to add financial climate risk in the context of the banking industry.

Financial risk management procedures are normally outlined in bank procedures manuals, which explain how to finance, mitigate, and generally manage the levels of risk (credit, liquidity, market and operational) assumed by a bank, as well as how to finance, mitigate, and manage them, based on a risk appetite that is compatible with the company's business model and its capacity to finance the risks. Physical and transition risks are a new type of risk in this regard, the current methodological framework it does not fully distinguish as the fitth and sixth risk or as inherent risk to each of the above four risk mentioned. In some cases, the study of companies is still a field that needs to be explored and researched at the macro level.

Usually, scientific research is focused on macro-level analysis (e.g. mortgage entities, banks or insurers), but at the micro level (retail for low-income households), there is a lack of assessments focused on analyzing micro-level data of companies considering financial climate approach. Within this particular context, the literature reviewed above on banks frameworks are primarily targeted to regulated lending firms, however, medium to small size lending firms, including small lending organizations, not necessarily are in the scope of Basel Accords have not yet been addressed by the existing models that have been established. This presents a promising area for further research.

V.  References

Allen Steven. (2013). *Financial Risk Management*. John Wiley & Sons.

Allen, T., Dees, S., Boissinot, J., Mateo, C., Graciano, C., Chouard, V., Clerc, L., De Gaye, A., Devulder, A., Diot, S., Lisack, N., Pegoraro, F., Rabaté, M., Svartzman, R., & Vernet, L. (2020). *Climate-Related Scenarios for Financial Stability Assessment: an Application to France*.

Alogoskoufis, S., Dunz, N., Emambakhsh, T., Hennig, T., Kaijser, M., Kouratzoglou, C., Muñoz, M. A., Parisi, L., & Salleo, C. (2021). *Occasional Paper Series ECB economy-wide climate stress test Methodology and results*. https://doi.org/10.2866/460490

Andersson, M., Bolton, P., & Samama, F. (2016). Hedging climate risk. In *Financial Analysts Journal* (Vol. 72, Issue 3, pp. 13–32). CFA Institute. https://doi.org/10.2469/faj.v72.n3.4

AON. (2024). *The value of CAT models for measuring climate risk*. https://www.aon.com/apac/insights/blog/the-value-of-cat-models-for-measuring-climate-risk

Apergis, N., Poufinas, T., & Antonopoulos, A. (2022). ESG scores and cost of debt. *Energy Economics*, *112*. https://doi.org/10.1016/j.eneco.2022.106186

Apostolik, R., Donohue, C., & Went, P. (2009). *An Overview of Banking, Banking Risks, and Risk-Based Banking Regulation Foundations of Banking Risk*.

ASB. (2021). *Catastrophe Modeling (for All Practice Areas)*.

Baesens, B., & van Gestel, T. (2009a). *Credit Risk Management Basic Concepts: financial risk components, rating analysis, models, economic and regulatory capital*.

Baesens, B., & van Gestel, T. (2009b). *Credit Risk Management Basic Concepts: financial risk components, rating analysis, models, economic and regulatory capital*.

Bank of England. (2019). *Financial Policy Committee Prudential Regulation Committee The 2021 biennial exploratory scenario on the financial risks from climate change*. www.bankofengland.co.uk/legal/privacy.

Banque of France. (2020). *A first assessment of financial risks stemming from climate change*.

Batten, S., Sowerbutts, R., & Tanaka, M. (2016). *Staff Working Paper No. 603 Let's talk about the weather: the impact of climate change on central banks*. www.bankofengland.co.uk/research/Pages/workingpapers/default.aspx

Battiston, S. (2022). The NACE-CPRS-IAM mapping: A tool to support climate risk analysis of financial portfolio using NGFS scenarios. *SSRN*. https://ec.europa.eu/eurostat/statistics-

Battiston, S., Mandel, A., Monasterolo, I., Schütze, F., & Visentin, G. (2017). A climate stress-test of the financial system. *Nature Climate Change*, *7*(4), 283–288. https://doi.org/10.1038/nclimate3255

Bennani Leila, Guenedal Theo Le, Lepetit Frederic, Ly Lai, Mortier Vincent, & Sekine Takaya. (2018). *The Alpha and Beta of ESG investing*.

Bhinderwala Shiraj. (1995). *Insurance Loss Analysis of Single Family Dwellings Damaged in Hurricane Andrew*. Clemson University.

Bikakis, A. (2020). *Climate Change, Flood Risk and Mortgages in the UK: a Scenario Analysis*.

BIS. (2024a). *Basel Committee on Banking Supervision*. www.bis.org

BIS. (2024b). *Basel Committee on Banking Supervision The Basel Framework*.

Bolton, P., & Kacperczyk, M. (2023). Global Pricing of Carbon-Transition Risk. *Journal of Finance*, *78*(6), 3677–3754. https://doi.org/10.1111/jofi.13272

Buckley, N., Hamilton, A., Harding, J., Roche, N., Ross, N., Sands, E., Skelding, R., Watford, N., & Whitlow, H. (2001). *EUROPEAN WEATHER DERIVATIVES WORKING PARTY MEMBERS*.

Byers, H. R., Landsberg, H. E., Wexler, H., Haurwitz, B., Spilhaus, A. F., Willett, H. C., & Houghton, H. G. (1951). *Compendium of Meteorology* (T. F. Malone, Ed.). American Meteorological Society. https://doi.org/10.1007/978-1-940033-70-9

Calabrese, R., Dombrowski, T., Mandel, A., Pace, R. K., & Zanin, L. (2022). *Impacts of extreme weather events on mortgage risks and their evolution under climate change: A case study on Florida*. https://ssrn.com/abstract=3929927

Caldecott, B., Harnett, E., Cojoianu, T., Kok, I., Pfeiffer, A., & Rios, A. R. (2016). *Stranded Assets: A Climate Risk Challenge*. http://myidb.iadb.org/publications/resources/img/by-nc-nd.png

Caldecott, B., Howarth, N., & Mcsharry, P. (2013). *Stranded Assets in Agriculture: Protecting Value from Environment-Related Risks*.

Caldecott, B., & Smith School of Enterprise and the Environment. (2016). *Stranded assets and thermal coal : an analysis of environment-related risk exposure*.

Caloia, F., & Jansen, D.-J. (2021). *Flood risk and financial stability: Evidence from a stress test for the Netherlands*.

Carney Mark. (2015). *Breaking the Tragedy of the Horizon-climate change and financial stability*.

CAS. (2020). *Exceedance Probability in Catastrophe Modeling*.

Clark Karen. (1986). A Formal Approach to Catastrophe Risk Assessment and Management. *Proceedings of the CasualtyActuarial Society* , *73*(140).

Colas J, Khaykin I, Pyanet A, & Westheim J. (2018). *Extending Our Horizons*.

Committee Expert Group on Climate Change, S., & Group on Securities Statistics, W. (2024). *Statistics Paper Series Climate change-related statistical indicators Statistics Committee Expert Group on*

*Climate Change and Statistics and Working Group on Securities Statistics No 48*.
https://doi.org/10.2866/059096

Considine, G. (2000). *Introduction to Weather Derivatives*. http://www.cme.com/weather/index.html

Contat Carrie Hopkins Luis Mejia Matthew Suandi, J., Contat, J., Hopkins, C., Mejia, L., Suandi, M.,
Contat Carrie Hopkins, J., & Luis Mejia Matthew Suandi, fhfagov. (2023). *When Climate Meets Real
Estate: A Survey of the Literature*. https://www.fhfa.gov/papers/wp2305.aspx.

Crompton, R. P., & McAneney, K. J. (2008). Normalised Australian insured losses from meteorological
hazards: 1967–2006. *Environmental Science & Policy*, *11*(5), 371–378.
https://doi.org/10.1016/j.envsci.2008.01.005

Culver, C. G., Lew, H. S., Hart, G. C., & Pinkham, C. W. (1975). *Natural hazards evaluation of existing
buildings*. https://doi.org/10.6028/NBS.BSS.61

De Bandt, O., Kuntz, L.-C., Pankratz, N., Pegoraro, F., Solheim, H., Sutton, G., Takeyama, A., & Xia, D.
(2023). *Basel Committee on Banking Supervision The effects of climate change-related risks on
banks: a literature review*. www.bis.org/bcbs/

Drei, A., Le Guenedal, T., Lepetit, F., Mortier, V., Roncalli, T., & Sekine, T. (2020). ESG Investing in Recent
Years: New Insights from Old Challenges. *SSRN Electronic Journal*.
https://doi.org/10.2139/ssrn.3683469

DTCC. (2023). *Climate-Related Financial Risk*.

Duan, T., Li, F. W., Wen, Q., Gomes, F., Hsu, P.-H., Kacperczyk, M., Koelbel, J. F., Lee, H., Li, K., Loh, R.,
Matos, P., Mukherjee, A., Nozawa, Y., Taylor, L., Sulaeman, J., Wang, T., Yao, C., Zhang, B., Zhang,
C., ... Zhu, Q. (2021). *Is Carbon Risk Priced in the Cross-Section of Corporate Bond Returns?*
https://www.bankofengland.co.uk/news/2020/july/statement-on-banks-commitment-to-

Dunz, N., Naqvi, A., & Monasterolo, I. (2021). Climate sentiments, transition risk, and financial stability in
a stock-flow consistent model. *Journal of Financial Stability*, *54*.
https://doi.org/10.1016/j.jfs.2021.100872

EIOPA. (2017). *Open-source tools for the modelling and management of climate change risks-European
Union*. https://www.gnu.org/licenses/gpl-3.0.html

Eliwa, Y., Aboud, A., & Saleh, A. (2021). ESG practices and the cost of debt: Evidence from EU countries.
*Critical Perspectives on Accounting*, *79*. https://doi.org/10.1016/j.cpa.2019.102097

Engle, R. F., Giglio, S., Kelly, B., Lee, H., & Stroebel, J. (2020). Hedging climate change news. *Review of
Financial Studies*, *33*(3), 1184–1216. https://doi.org/10.1093/rfs/hhz072

Faccini, R., Matin, R., Skiadopoulos, G., Albuquerque, R., Branikas, I., Hiraki, K., Fasois, C. L.,
Malliaropoulos, D., Mirone, G., Mølbak, M., Kacperczyk, M., Konstan-Tinidi, E., Kostakis, A.,
Papakonstantinou, F., Papanikolaou, D., Ramelli, S., Rognone, L., Sautner, Z., Spyridopoulos, I., ...

Tsou, C.-Y. (2023). *Dissecting Climate Risks: Are they Reflected in Stock Prices? *.* https://ssrn.com/abstract=3795964

Fama, E. F. (1970a). Efficient Capital Markets: A Review of Theory and Empirical Work. *The Journal of Finance*. https://about.jstor.org/terms

Fama, E. F. (1970b). Efficient Capital Markets: A Review of Theory and Empirical Work. In *Source: The Journal of Finance* (Vol. 25, Issue 2).

Fiedler, T., Pitman, A. J., Mackenzie, K., Wood, N., Jakob, C., & Perkins-Kirkpatrick, S. E. (2021). Business risk and the emergence of climate analytics. *Nature Climate Change*, *11*(2), 87–94. https://doi.org/10.1038/s41558-020-00984-6

Giuzio, M., Krusec, D., Levels, A., Melo, A. S., Mikkonen, K., & Radulova, P. (2019). *Climate change and financial stability*. https://www.ecb.europa.eu/press/financial-stability-publications/fsr/special/html/ecb.fsrart201905_1~47cf778cc1.en.html

Gordy, M. B. (2000). A comparative anatomy of credit risk models. *Journal of Banking & Finance*. www.elsevier.com/locate/econbase

Gourdel, R., Sydow, M., & Central Bank, E. (2021). *Bi-layer stress contagion across investment funds: a climate application*.

Green Bonds, U. S., Baker, M., Bergstresser, D., Serafeim, G., Wurgler, J., Library, B., & Hall, M. (2018). *Financing the Response to Climate Change: The Pricing and Ownership of*. https://climate.nasa.gov/evidence/.

Grippa, P., Mann, S., Čihák, M., Hofman, D. J V, & Qureshi, M. S. (2020). *Climate-Related Stress Testing: Transition Risks in Norway, WP/20/232, November 2020*.

Grossi Patricia, & Kunreuther Howard. (2005). *Catastrophe Modeling: A New Approach to Managing Risk*. Springer.

Gurvich, A., & Creamer, G. G. (2022). Carbon Risk Factor Framework. *The Journal of Portfolio Management*, *48*(10), 148–164. https://doi.org/10.3905/jpm.2022.1.416

Harris, J. (2015). *Special Report The Carbon Risk Factor (EMI-'Efficient Minus Intensive')*.

Hoepner, A. G. F., Oikonomou, I., Sautner, Z., Starks, L. T., & Zhou, X. Y. (2023). *ESG Shareholder Engagement and Downside Risk*. http://ssrn.com/abstract_id=2874252www.ecgi.global/content/working-papers

Howard A, & Patrascu O. (2017). *Primer: building a case for infrastructure finance Climate change: redefining the risks*. http://www.mckinsey.com/business-functions/sustainability-and-resource-productivity/our-

Huang, Z., Rosowsky, D. V., & Sparks, P. R. (2001). Long-term hurricane risk assessment and expected damage to residential structures. *Reliability Engineering & System Safety*, *74*(3), 239–249. https://doi.org/10.1016/S0951-8320(01)00086-2

Hurd Tom. (2010). *Chapter 4 Structural Models of Credit Risk: Mimeo*.

IFRS. (2023). *Climate-related Disclosures IFRS S2 IFRS ® Sustainability Disclosure Standard International Sustainability Standards Board*.

Ishtiaq, M. (2015). *Risk Management in Banks: Determination of Practices and Relationship with Performance*. UNIVERSITY OF BEDFORDSHIRE.

Jewson, S. (2004). Introduction to Weather Derivative Pricing. *The Journal of Alternative Investments*. www.evomarkets.com

Le Guenedal, T. (2022). Financial Modeling of Climate-related Risks. *Hal Open Science.* https://theses.hal.science/tel-04013805

Mccarthy, R., Reid, G., & Silva -Getty Images, J. (2024). *Assessing Physical Climate Risk in Repo-eligible Residential Mortgage-backed Securities*.

Merton, R. C. (1974). On the Pricing of Corporate Debt: The Risk Structure of Interest Rates ON THE PRICING OF CORPORATE DEBT: THE RISK STRUCTURE OF INTEREST RATES*. In *Source: The Journal of Finance* (Vol. 29, Issue 2).

Monnin, P. (2018). *Integrating Climate Risks into Credit Risk Assessment Current Methodologies and the Case of Central Banks Corporate Bond Purchases*. https://ssrn.com/abstract=3350918

Naqvi, A., & Monasterolo, I. (2021). Assessing the cascading impacts of natural disasters in a multi-layer behavioral network framework. *Scientific Reports*, *11*(1). https://doi.org/10.1038/s41598-021-99343-4

Nguyen, Q., Diaz-Rainey, I., Kuruppuarachchi, D., Mccarten, M., & Tan, E. K. M. (2020). *Climate Transition Risk in U.S. Loan Portfolios: Are All Banks The Same?* https://ssrn.com/abstract=3766592

Nordhaus, W. D. (1993). *Optimal Greenhouse-Gas Reductions and Tax Policy in the "DICE" Model* (Vol. 83, Issue 2). American Economic Association.

of Japan, B. (2022). *Physical risks from climate change faced by Japan's financial institutions: Impact of floods on real economy, land prices, and FIs' financial conditions*. www.emdat.be

Paltsev, S., Reilly, J. M., Jacoby, H. D., Eckaus, R. S., Mcfarland, J., Sarofim, M., Asadoorian, M., Babiker, M., & Prinn, R. G. (2005). *MIT Joint Program on the Science and Policy of Global Change The MIT Emissions Prediction and Policy Analysis (EPPA) Model: Version 4 The MIT Emissions Prediction and Policy Analysis (EPPA) Model: Version 4*.

Pinchot, A., Zhou, L., Christianson, G., McClamrock, J., & Sato, I. (2021). Assessing Physical Risks from Climate Change: Do Companies and Financial Organizations Have Sufficient Guidance? *World Resources Institute*. https://doi.org/10.46830/wriwp.19.00125

Pita, G., Pinelli, J.-P., Gurley, K., & Mitrani-Reiser, J. (2015). State of the Art of Hurricane Vulnerability Estimation Methods: A Review. *Natural Hazards Review*, *16*(2). https://doi.org/10.1061/(asce)nh.1527-6996.0000153

Raimo, N., Caragnano, A., Zito, M., Vitolla, F., & Mariani, M. (2021). Extending the benefits of ESG disclosure: The effect on the cost of debt financing. *Corporate Social Responsibility and Environmental Management*, *28*(4), 1412–1421. https://doi.org/10.1002/csr.2134

Riordan, R., Jacob A, Nerlinger M, Riordan R, Rohleder M, & Wilkens M. (2020). *Carbon Risk\**. https://ssrn.com/abstract=2930897

Roncalli, T., Le Guenedal, T., Lepetit, F., Roncalli, T., & Sekine, T. (2021). *The Market Measure of Carbon Risk and its Impact on the Minimum Variance Portfolio \**. https://carima-project.de/en/downloads

Roncoroni, A., Battiston, S., Onésimo, L., Farfán, L. E., & Jaramillo, S. M. (2021). *Climate risk and financial stability in the network of banks and investment funds \**.

Sassi, O., Crassous, R., Hourcade, J. C., Gitz, V., Waisman, H., & Guivarch, C. (2010). IMACLIM-R: A modelling framework to simulate sustainable development pathways. *International Journal of Global Environmental Issues*, *10*(1–2), 5–24. https://doi.org/10.1504/IJGENVI.2010.030566

Sciaudone J, Feuerborn D, Rao G, & Daneshvaran S. (1997). De- velopment of objective wind damage functions to predict wind damage to low-rise structures. *U.S. National Conf. on Wind Engineering*.

Sill, B. L., & Kozlowski, R. T. (1997). Analysis of Storm-Damage Factors for Low-Rise Structures. *Journal of Performance of Constructed Facilities*, *11*(4), 168–177. https://doi.org/10.1061/(ASCE)0887-3828(1997)11:4(168)

Skoglund Jimmy, & Chen Wei. (2015). *Financial Risk Management*. John Wiley & Sons.

Slimane, M. Ben, Le Guenedal, T., & Roncalli, T. (2019). *ESG Investing in Corporate Bonds: Mind the Gap \**. https://ssrn.com/abstract=3683472

Sparks, P. R., Schiff, S. D., & Reinhold, T. A. (1994). Wind damage to envelopes of houses and consequent insurance losses. *Journal of Wind Engineering and Industrial Aerodynamics*, *53*(1–2), 145–155. https://doi.org/10.1016/0167-6105(94)90023-X

Stern, N. (2008). The Economics of Climate Change. In *Source: The American Economic Review* (Vol. 98, Issue 2).

Stewart, M. G. (2003). Cyclone damage and temporal changes to building vulnerability and economic risks for residential construction. *Journal of Wind Engineering and Industrial Aerodynamics*, *91*(5), 671–691. https://doi.org/10.1016/S0167-6105(02)00462-2

Szulczyk Kenneth. (2014). *Money, Banking, and International Finance* (Edition 2). Kenneth R. Szulczyk.

TCFD. (2017). *Recommendations of the Task Force on Climate-related Financial Disclosures i Letter from Michael R. Bloomberg*.

TCFD. (2023). *Task Force on Climate-related Financial Disclosures 2023 Status Report*.

Thom J, & Ralite S. (2019). *Storm ahead, a proposal for a climate stress-test scenario*.

Turner, J. (2023). Climate Change Physical Risk in Catastrophe Modelling. *Journal of Catastrophe Risk and Resilience*. https://journalofcrr.com/climate-change-physical-risk-in-catastrophe-modelling/#:~:text=Climate

Unanwa, C. O., McDonald, J. R., Mehta, K. C., & Smith, D. A. (2000). The development of wind damage bands for buildings. *Journal of Wind Engineering and Industrial Aerodynamics*, *84*(1), 119–149. https://doi.org/10.1016/S0167-6105(99)00047-1

Van-Deventer Donarld R., Imai Kenji, & Mesler Mark. (2013). *Advanced Financial Risk Management*. John Wiley & Sons Singapore.

Vermeulen, R., Schets, E., Lohuis, M., Kölbl, B., Jansen, D.-J., & Heeringa, W. (2018). *An energy transition risk stress test for the financial system of the Netherlands*. www.dnb.nl

Vermeulen, R., Schets, E., Lohuis, M., Kölbl, B., Jansen, D.-J., & Heeringa, W. (2019). *The Heat is on: A framework measuring financial stress under disruptive energy transition scenarios*. https://ssrn.com/abstract=3346466

Vickery, P. J., Skerlj, P. F., Lin, J., Twisdale, L. A., Young, M. A., & Lavelle, F. M. (2006). HAZUS-MH Hurricane Model Methodology. II: Damage and Loss Estimation. *Natural Hazards Review*, *7*(2), 94–103. https://doi.org/10.1061/(ASCE)1527-6988(2006)7:2(94)

Walker George. (2011). Modelling the vulnerability of buildings to wind — a review. *Canadian Journal of Civil Engineering*, *38*.

Weatherwise. (1954). The Travelers Weather Research Center. *Weatherwise*, *7*(6), 159–159. https://doi.org/10.1080/00431672.1954.9930352

Weyzig, F., Kuepper, B., Willem Van Gelder, J., & Van Tilburg, R. (2014). *The Price of Doing Too Little Too Late Green New Deal Series volume 11*. www.gef.eu

Zerbib, O. D. (2019). The effect of pro-environmental preferences on bond prices: Evidence from green bonds. *Journal of Banking and Finance*, *98*, 39–60. https://doi.org/10.1016/j.jbankfin.2018.10.012